\documentclass[twocolumn]{aastex631}

\usepackage{graphicx}	% Including figure files
\usepackage{amsmath}	% Advanced maths commands
\usepackage{ulem}
\usepackage[version=4]{mhchem}
\usepackage[figuresright]{rotating}
\usepackage[percent]{overpic}
\usepackage{float}
\usepackage[T1]{fontenc}
\usepackage{tabularx}
\usepackage{enumitem}
\usepackage{soul}

\shorttitle{Chemical Modeling of \ce{C2H4O2} isomers in NGC 6334I}
\shortauthors{Shope et al.}

\begin{document}

\title[Modeling \ce{C2H4O2} isomer ratios in NGC 6334I]
{Interstellar Glycolaldehyde, Methyl Formate, and Acetic Acid. II. Chemical Modeling of the Bimodal Abundance Pattern in NGC 6334I}

\correspondingauthor{Brielle Shope}
\email{bmt5hv@virginia.edu}

\author[0000-0003-4147-4125]{Brielle M. Shope}
\affiliation{Department of Chemistry, University of Virginia, Charlottesville, VA 22904, USA}

\author[0000-0002-1577-5322]{Samer J. El-Abd}
\affiliation{Department of Astronomy, University of Virginia, Charlottesville, VA 22904, USA}

\author[0000-0002-6558-7653]{Crystal L. Brogan}
\affiliation{National Radio Astronomy Observatory, Charlottesville, VA 22903, USA}

\author[0000-0001-6492-0090]{Todd R. Hunter}
\affiliation{National Radio Astronomy Observatory, Charlottesville, VA 22903, USA}

\author[0000-0002-7475-3908]{Eric R. Willis}
\affiliation{Department of Chemistry, University of Virginia, Charlottesville, VA 22904, USA}

\author[0000-0003-1254-4817]{Brett A. McGuire}
\affiliation{Department of Chemistry, Massachusetts Institute of Technology, Cambridge, MA 02139, USA}
\affiliation{National Radio Astronomy Observatory, Charlottesville, VA 22903, USA}

\author[0000-0001-7723-8955]{Robin T. Garrod}
\affiliation{Department of Chemistry, University of Virginia, Charlottesville, VA 22904, USA}
\affiliation{Department of Astronomy, University of Virginia, Charlottesville, VA 22904, USA}

\begin{abstract}
Gas-phase abundance ratios between \ce{C2H4O2} isomers methyl formate (MF), glycolaldehyde (GA), and acetic acid (AA) are typically on the order of 100:10:1 in star-forming regions. However, an unexplained divergence from this neat relationship was recently observed towards a collection of sources in the massive protocluster NGC 6334I; some sources exhibited extreme MF:GA ratios, producing a bimodal behavior between different sources, while the MF:AA ratio remained stable. Here, we use a three-phase gas-grain hot-core chemical model to study the effects of a large parameter space on the simulated \ce{C2H4O2} abundances. A combination of high gas densities and long timescales during ice-mantle desorption ($\sim$125--160~K) appears to be the physical cause of the high MF:GA ratios. The main chemical mechanism for GA destruction occurring under these conditions is the rapid adsorption and reaction of atomic H with GA on the ice surfaces before it has time to desorb. The different binding energies of MF and GA on water ice are crucial to the selectivity of the surface destruction mechanism; individual MF molecules rapidly escape the surface when exposed by water loss, while GA lingers and is destroyed by H. Moderately elevated cosmic-ray ionization rates can increase absolute levels of COM production in the ices and increase the MF:GA ratio, but extreme values are destructive for gas-phase COMs. We speculate that the high densities required for extreme MF:GA ratios could be evidence of COM emission dominated by COMs desorbing within a circumstellar disk.
\end{abstract}

\keywords{astrochemistry --- stars: formation --- ISM: abundances --- ISM: molecules --- dust}

\section{Introduction} \label{sec:intro}
Methyl formate (MF, \ce{HCOOCH3}), glycolaldehyde (GA, \ce{CH2(OH)CHO}), and acetic acid (AA, \ce{CH3COOH}) -- all structural isomers of formula \ce{C2H4O2} -- are commonly detected in the gas phase toward hot molecular cores and their low-mass analogs, hot corinos, through their cm, mm, and sub-mm wavelength rotational emission \citep{Brown_1975, Cazaux_2003, Hollis_2000,Jorgensen_2012, Mehringer_1997, jorg2016}. 
In an interstellar context, these three species are known as ``complex organic molecules'' (COMs), which are defined as carbon-bearing molecules composed of 6 or more atoms \citep{herbst_2009}. As with many other COMs, they are of some prebiotic interest; glycolaldehyde is a sugar-like diose and may be a key RNA precursor \citep{Benner2019}. Methyl formate leads to the synthesis of bio-polymers \citep{Occhiogrosso2011}, while acetic acid is a metabolic intermediate, occurring naturally in body fluids.

Although some COMs may have efficient gas-phase routes to their formation in hot cores \citep[e.g. dimethyl ether, \ce{CH3OCH3};][]{garrod_herbst_2006}, COMs in general are thought to be formed initially on dust-grain surfaces at relatively low temperatures \citep{Garrod_2008,fedoseev_2015,Ioppolo21}, and then released into the gas phase as the dust grains are heated by the protostar. Shock-related heating and/or sputtering could also be important in some instances \citep{jorg_review2020}. Recent observations toward the quiescent Galactic Center cloud G+0.693$-$0.027 have also revealed a rich COM chemistry related to shock-induced release of dust-grain ice mantles \citep{zeng2018,zeng2020,rivilla2022a}.

The precise mechanisms leading to the production of each individual COM molecule remain under debate, but much of the chemical modeling work on hot cores and corinos over the past decade or so has concentrated on the recombination of functional-group radicals on the dust grains or within their ice mantles \citep{garrod_herbst_2006,Garrod_2008, Garrod_2013}. As some of these radicals, such as \ce{CH3O} and \ce{CH2OH}, may derive from the same source molecules, i.e.~CO, \ce{H2CO} (formaldehyde) and/or \ce{CH3OH} (methanol), the observational study of structural isomers (e.g. MF and GA) that contain these functional groups may provide constraints on astrochemical models of star-formation chemistry and the underlying chemical mechanisms that they simulate.

Glycolaldehyde was first detected in the ISM toward the high-mass star-forming region Sgr B2(N) by \citet{Hollis_2000}, who derived ratios for MF:GA:AA of 26:4:1 based on observations with the NRAO 12~m telescope. Follow-up mapping work by \citet{Hollis_2001}, using the BIMA interferometer, revealed GA to be highly spatially extended in that source (contrary to the compact emission of MF and AA). Those authors determined MF:GA:AA ratios of 52:1:2 specifically toward the Sgr B2(N) core. Detection of GA toward this source was again confirmed by \citet{Halfen_2006}. The more recent EMoCA survey of Sgr B2(N) by \citet{belloche_2016}, using ALMA, yielded ratios 60:7:1 (based on their upper limit for AA) for hot core Sgr B2(N2).

The first detection of glycolaldehyde toward a solar-type protostar \citep[IRAS 16293-2422B;][]{Jorgensen_2012}, obtained with ALMA, indicated a ratio of MF to GA of $\sim$10-15. The later ALMA PILS line survey of the same source yielded MF:GA:AA ratios of 93:12:1 \citep{jorg2016}. The EMoCA ALMA line survey of Sgr B2(N2) yielded a MF:GA:AA ratio of 60:7:1 \citep{belloche_2016} and the ALMA observations of NGC 6334I MM1 sources yielded ratios of 130-380:10:1 \citep{El_Abd_2019}.

Based on these and other observations of \ce{C2H4O2} isomers toward both high- and low-mass star-forming cores, approximate MF:GA:AA ratios of 100:10:1 have been considered broadly representative of these species' relative abundances. However, a divergence from this neat relationship -- particularly between MF and GA -- was revealed by the study of \citet{El_Abd_2019}, who examined the abundance ratios of the \ce{C2H4O2} isomers toward a collection of sources within the massive protocluster NGC6334I, along with values for a variety of other sources taken from the literature. The column densities of MF maintained stable ratios against AA amongst all the sources (see their Figure 2A). However, when comparing the column densities of MF with those of GA, a bimodal distribution was observed among a number of sources, including locations within NGC6334I MM2 (their Figure 2B). Along with some more typical MF:GA ratios of $\sim$10:1, some of these high-mass cores had MF:GA ratios that were at least an order of magnitude higher. \citet{El_Abd_2019} speculated that this bimodal behavior may indicate some unknown competing pathways and/or unknown physical processes in the protostellar environment that must be taken into account in the models in order to explain the observations.

Simulating the observed MF:GA:AA ratios using astrochemical models has been a challenge in the past, due to the models predicting excessive amounts of glycolaldehyde \citep{Garrod_2008, Garrod_2013}. In these past treatments, production of each of the three isomers depended mainly upon diffusive reactions between radicals on grain surfaces, occurring at somewhat elevated temperatures \citep[$\sim$20--40~K;][]{Garrod_2008} during the warm-up of the protostellar envelope. More recently, \citet{Jin_2020} and \citet[][G22 hereafter]{Garrod_2022} made substantial changes to their astrochemical models, to include nondiffusive grain-surface and bulk-ice chemistry, whereby reactants may be brought together not as the result of diffusion but through some other process, such as a preceding reaction. The earlier \citet{Garrod_2011} model included a formulation for nondiffusive \ce{CO2} production through the facile process of atomic H diffusion to meet atomic O and instantly react with nearby surface CO.

This type of nondiffusive reaction mechanism, labeled more specifically as a ``three-body'' (3-B) reaction by \citet{Jin_2020}, was applied to a selection of reactions leading to COM production. Also included was the ``photodissociation-induced'' (PDI) reaction process, based on a similar treatment used for comet ice chemistry by \citet{Garrod_2019}, in which a nondiffusive meeting would occur as the result of the spontaneous production of one reactant in the presence of the other via the photodissociation of some precursor molecule near to one of the reactants. \citet{Jin_2020} also included the Eley-Rideal (E-R) mechanism as a related nondiffusive process, along with a special case of the 3-B process in which the formation energy of the reaction product of the initiating reaction could allow it to overcome the activation energy of the subsequent reaction (known as ``three-body excited formation'', 3-BEF). The G22 model included the same set of nondiffusive mechanisms, applying them comprehensively to the entire network of surface and bulk-ice reactions.

The inclusion of the 3-B and PDI mechanisms in particular allowed various COMs to be formed effectively in the dust-grain ice mantles or on the grain surfaces, without the need for a direct diffusion mechanism to mediate the reactions. The PDI process correctly allows for photochemistry to occur in the ices even at very low temperatures \citep[see, e.g.,][]{oberg_2009}. Furthermore, the 3-B process allows for COMs to be formed as the result of the surface chemistry that builds up the ice mantles under cold and dark cloud conditions, driven mainly by the diffusion of atomic H.

Perhaps most importantly for the \ce{C2H4O2} isomers, the formation of MF through the 3-B process, as a side-effect of grain-surface methanol production, avoids the problematic \ce{CH3O}:\ce{CH2OH} photodissociation branching ratio of 1:5 \citep{oberg_2009}. Both of these radicals are instead formed either through hydrogenation of formaldehyde (\ce{H2CO}) or by H-abstraction from methanol by H atoms. The latter process, which favors \ce{CH2OH} production, is relatively slow, whereas the former process is an essential step in methanol production that strongly favors the production of \ce{CH3O}. The models of G22 included further reaction pathways for the formation of glycolaldehyde from glyoxal (HCOCHO), as well as interconversion between glyoxal, glycolaldehyde, and ethylene glycol (\ce{(CH2OH)2)}, through barrier-mediated H-addition and abstraction reactions with atomic H, using activation barriers partially based on rates provided by \citet{simons_2020}, who based their rate parameters on calculations by \citet{barcia_2018}. These additional pathways were found to contribute significantly to the total production of glycolaldehyde.

As with past hot-core chemistry models by those authors, G22 used a physical treatment deemed to be representative of the evolution of a generic, star-forming source. The treatment divides the evolution into two stages; the first stage involves a cold, free-fall collapse to the final density of the hot core, during which the ice mantles form. In the second, warm-up stage, the gas and dust temperatures rise over time, allowing the ice mantles ultimately to be desorbed into the gas phase, typically at temperatures greater than around 100~K.

The timescale of warm-up has typically been the only free parameter considered in these models. However, \citet{Barger_2020} conducted a more substantial survey of the parameter space of the physical treatment, using a model that did not include nondiffusive grain-surface/ice chemistry. They found that the combination of the chosen warm-up timescale and cosmic-ray ionization rate (CRIR) could produce broad, and somewhat degenerate variation in gas-phase and grain-surface abundances for COMs.

CRIR values are rather poorly constrained in the ISM in general, with recently determined values diverging from the canonical value of $\zeta \simeq 10^{-17}$~s$^{-1}$, reaching as many as three orders of magnitude higher, dependent on source \citep[e.g.][]{Caselli98,vanderTak_2006,favre_2018}. Theoretical work indicates that the CRIR from Galactic cosmic rays should be dependent on H$_2$ column density \citep[e.g.][]{Padovani_2009,Rimmer_2012,Padovani_2018,FitzAxen_2021}. More recent theoretical and observational work provides evidence that high-energy, ionizing particles may be produced locally within protostellar sources \citep[e.g.][]{Padovani_2018,Cabedo_2023,Sabatini_2023}. Thus, the operative CRIR used in models of particular sources may require a far more localized tuning than the simple adoption of a generic CRIR value. 

Other physical conditions, such as the gas density and visual extinction behavior, may also influence the gas and grain chemistry in a meaningful way: the former determining the rate of growth of ice mantles and the rates of gas-phase chemistry; the latter determining the dust-grain temperature and the degree of photo-processing of the ice mantles to form or destroy COMs.

Here, we present hot-core chemical model results from a broad physical parameter space using the MAGICKAL chemical model as presented by G22, with a particular emphasis on the search for parameter values that lead to large MF:GA abundance ratios and the possible emergence of bimodal behavior between different models in the grid. The general effects of each physical parameter on the model are also considered. The main parameters explored within the context of the usual physical treatment are the peak (final) hydrogen number density ($n_{\text{H}}$), cosmic ray ionization rate ($\zeta$), and the warm-up timescale ($t_{\text{wu}}$) of the hot core. The influence of changing the initial visual extinction ($A_\text{{v,initial}}$) of the collapsing cloud is also tested.

By varying these parameters, we investigate which parameter combinations best reproduce the observed isomer abundance ratios in the twelve sources of the star forming regions MM1 and MM2 in the massive protocluster NGC6334I \citep{El_Abd_2019}. The results are also compared with molecular abundance ratios obtained by the PILS and EMoCA line surveys of IRAS 16293 and Sgr B2(N), respectively, presented in Appendix \ref{appendix}.

A summary of the general effects of each parameter on the model is given in Section~\ref{effects}.

%%%%%%%%%%%%%%%%%%%%%%%%%%%%%%%%%%%%%%%%%%%
\section{Methods}

The gas-grain chemical kinetics model MAGICKAL is used to simulate the time-dependent chemistry in hot cores. MAGICKAL \citep[{\it Model for Astrophysical Gas and Ice Chemical Kinetics And Layering};][]{Garrod_2013} is a three-phase model used to calculate and solve the coupled rate equations that govern the chemistry occurring in the gas, grain-surface, and bulk-ice phases. Except for the chemical network employed, the chemical model used here is identical to the ``{\tt final}'' model setup presented by G22. Notable features of that model include both diffusive and nondiffusive grain-surface and bulk-ice chemical reaction mechanisms. While all chemical species on the grains are in principle allowed to diffuse on the surface, diffusion within the bulk ice is limited to H and H$_2$, which are assumed to exist in interstitial positions within the ice matrix. Bulk-ice diffusion rates for H and H$_2$ are based on barriers twice the strength of the surface barriers. Tunneling rates are used (based on a barrier width of 3.2\AA, corresponding to a majority water ice) in cases where those rates exceed the thermal diffusion rates.  As per G22, the surface binding energy of atomic H is taken as 661~K and the diffusion barrier as 243~K, corresponding to the average values calculated by \cite{SENEVIRATHNE201759} for H on ASW. As in G22, the sticking coefficient for all species is unity. The reader should refer to G22 for detailed descriptions of all of the main features of the model.

The chemical network used here is that presented by \cite{G&H23} for their M4 setup; it is identical to that of G22, except that it includes some additional gas-phase proton-transfer reactions between the protonated forms of various COMs and certain other species that have proton affinities larger than that of ammonia. For the \ce{C2H4O2} isomers and related species, the network may be considered essentially identical to that of G22; the most important proton-transfer reactions for these species involve ammonia (NH$_3$) and methanol (CH$_3$OH), and these were already present in the G22 network.

The chemical network includes a total of 749 gas-phase species, and 318 (neutral) grain-surface species, with the same number of species also traced in the bulk ice. There is a total of 22,401 reactions and processes, including transfer of atoms/molecules between the surface and bulk ice. This number also includes surface/bulk reactions that are replicated between the diffusive and various nondiffusive meeting mechanisms; the number of unique reactions on the grains (regardless of meeting mechanism or surface/ice phase) is 1,279. From this uniform chemical treatment, a grid of chemical model runs is constructed, based on the variation of a selection of key physical parameters related to the dynamical and thermal evolution of a hot core.

\subsection{Physical model}\label{phys}

The physical model has two consecutive stages of evolution: the first stage represents the isothermal, free-fall collapse of a dense core from an initially diffuse/translucent state, with the gas temperature held constant at 10~K. The initial density for each stage-1 run is set to the same value for all model runs \citep[$n_{\mathrm{H,initial}} = 3000$~cm$^{-3}$; ][]{Garrod_2022}, while the final gas density achieved during the collapse is varied in the grid. As the density increases during the collapse, the visual extinction rises according to 
$A_\mathrm{V}=A_\mathrm{V,initial} (n_{\mathrm{H}}/n_{\mathrm{H,initial}})^{2/3}$, 
where $A_\mathrm{V,initial}$ assumes a value of either 2 or 3 mag \citep{Garrod_2022}. These initial extinction and density values are consistent with typical values cited for translucent and/or clump material in interstellar clouds \citep{Snow_2006,Bergin_2007}, within which a core would grow. The visual extinction is capped at a value of 500 mag., which is sufficiently large to exclude effectively all external UV-vis photons. The dust temperature in stage 1 varies as a function of visual extinction \citep{Garrod_2011}, falling from $\sim$16 -- 8 K or from $\sim$14.7 -- 8 K, for initial visual extinctions of 2 or 3 mag., respectively. The dust grains are assumed to have a uniform radius of 0.1~$\mu$m, following past models. Chemical evolution during the collapse stage continues until the final density is reached, which occurs after $\sim$0.95 Myr. Most of the initial dust-grain ice build-up occurs during the collapse stage.

Stage 2 of the physical model involves the gradual warm-up of the gas and dust, while all other physical quantities are held steady (using their final values from stage 1). The dust temperature initially rises from 8~K until it reaches the 10~K initial temperature of the gas; $T_\mathrm{gas}$ and $T_\mathrm{dust}$ then rise together, until they reach a final temperature of 400~K. The characteristic warm-up timescale, $t_\mathrm{wu}$, which is varied within the model grid, corresponds to the time taken to reach 200~K, following past models. Parameter values used in the grid of models are shown in Table~\ref{table:conds} and described below.

\subsection{Model Grid Parameters}
\label{parameter}

Based on the general modeling treatment described above, we build a grid of chemical models adopting a range of physical parameter values (Table~\ref{table:conds}), in order to gauge the response of the chemistry beyond the analysis of G22. The standard parameter values used by those authors are shown in bold in the table. Aside from visual extinction, parameter values used in the model grid are distributed logarithmically around the standard values.

In the stage-1 setups, we adopt final hydrogen number densities, $n_{\mathrm{H}}$, from $2\times10^6$ to $2\times10^{10}$~cm$^{-3}$, in order to capture a range of possible gas-density values; typical hot-core gas densities are greater than $10^7$ cm$^{-3}$ \citep{choudhury}, while the typical, representative values adopted in our past models have ranged from $2\times10^7$ -- $2\times10^8$~cm$^{-3}$. To provide some additional parameter coverage within the latter range, a logarithmically intermediate value of $6.32\times10^7$~cm$^{-3}$ was included in the grid. The chosen final density, which is achieved at the end of stage 1, carries over to the corresponding stage-2 model runs.

The initial visual extinction in stage 1 is set at either 2 or 3 mag. G22, who tested these two values within a large grid of chemical parameter variations, found that the choice of $A_\mathrm{V,initial}$ made only a modest difference to simple ice abundances. In order to restrict the potentially large parameter space of the model grid, we choose to test only those same two values. By the end of the collapse stage, each model reaches at least 150 mag. of extinction, meaning that the post-collapse $A_\mathrm{V}$ value has minuscule influence over the chemical rates. The choice of initial visual extinction therefore has no practical effect on the behavior of the stage-2 models other than through the chemical abundances that they inherit from the stage-1 runs.

We also vary the cosmic ray ionization rates (CRIR), as most COM production on the grains is influenced in some degree by cosmic-ray induced UV photodissociation, while much of the post-desorption gas-phase destruction of COMs involves ion-molecule chemistry. 
The past chemical models of \citet{Barger_2020} suggested that CRIR values greater than the canonical value (of order 10$^{-17}$ s$^{-1}$) were most appropriate when compared with observed COM abundances, although the models used to draw this conclusion did not include any non-diffusive grain chemistry. In the present model grid, the CRIR ranges from an order of magnitude below our basic value of $1.30\times10^{-17}$ s$^{-1}$, to two orders of magnitude greater, with values distributed logarithmically. Each chosen CRIR value is applied throughout both stage 1 and 2, and does not vary with time or column density in these models.

For the warm-up timescales used in stage 2, seven values are adopted, ranging from $2\times10^4$ to $2\times10^6$ yr, based around the {\it medium} warm-up timescale used by G22 and others. Here we choose not to use the {\it fast} ($5\times10^4$ yr) and {\it slow} ($1\times10^6$ yr) values adopted in past models, as they fall between the logarithmically distributed values that we have adopted to ensure comparability within the grid. They are, however, within the extremes of the range of values tested here.

The full grid of physical parameters is varied to test every combination; this results in a total of 84 stage-1 models. For each stage-1 run, nine different stage-2 runs are calculated (varying warm-up timescale), such that the initial chemical and physical state of each of those nine is inherited from the same stage-1 run. The total number of stage-2 model runs is thus 756.

\begin{table}
\centering
\caption{Key physical parameters varied in the chemical model grid: initial visual extinction ($A_\text{v,init}$), hydrogen number density ($n_{\text{H}}$), cosmic ray ionization rate ($\zeta$), and warm-up timescale ($t_{\text{wu}}$). The table also indicates whether the parameter directly influences the physical behavior of the Stage-1 model, the Stage-2 model, or both. ``Standard'' values used in the \citet{Garrod_2022} models are shown in bold. The standard warm-up timescale indicated corresponds to the \textit{medium} value used in past models.\label{table:conds}}
\begin{tabular}{m{12mm} m{20mm} m{22mm} m{15mm}}
\hline
$A_{\text{v,init}}$ & n$_\text{H,final}$ & $\zeta$      & $t_{\text{wu}}$  \\
(mag)               & (cm$^{-3}$)        & (s$^{-1}$)   &(yr)  \\
\hline
{Stage 1}           &{Stages 1 \& 2}     &Stages 1 \& 2 &Stage 2 \\
\hline
	2	&	$2.00\times10^{6}$	        &	$1.30\times10^{-18}$	       & $2.00\times10^4$  \\
{\bf 3} &	$2.00\times10^{7}$	        &	$4.11\times10^{-18}$	       & $3.56\times10^4$  \\
		&	$6.32\times10^{7}$	        & {\bf 1.30$\times$10$^{-17}$} & $6.32\times10^4$  \\
	    & {\bf 2.00$\times$10$^{8}$}    &	$4.11\times10^{-17}$	       & $1.12\times10^5$  \\
		&	$2.00\times10^{9}$	        &	$1.30\times10^{-16}$	       & {\bf 2.00$\times$10$^5$}  \\
		&	$2.00\times10^{10}$	        &	$4.11\times10^{-16}$	       & $3.56\times10^5$  \\
		&						        &	$1.30\times10^{-15}$	       & $6.32\times10^5$  \\
		&						        &						           & $1.12\times10^6$  \\
		&						        &						           & $2.00\times10^6$  \\
\hline
\end{tabular}
\end{table}
%%%%%%%%%%%%%%%%%%%%%%%%%%%%%%%%%%%%%
%%%%%%%%%%%%%%%%%%%%%%%%%%%%%%%%%%%%%

\subsection{Observational Comparisons}
\label{sec:obs}

Quantitative comparisons of the model results with observational values can be used to determine which model best matches a particular source, thus indicating which physical conditions are most applicable in the context of our chemical model grid. The calculation of a matching parameter between a particular set of observational data and each chemical model may thus help to guide the explanation for the distinct chemical behavior observed between different sources.

Of particular interest is to identify any physical/chemical origins for the bimodal behavior seen in the MF and GA column densities observed by \citet{El_Abd_2019} in data from two Band 7 ALMA tunings toward the protocluster NGC 6334I. In their work, the column densities of MF, GA, and AA were obtained for lines of sight within the two sources in NGC 6334I, MM1 and MM2. Spectra were extracted from nine positions in MM1 and from three positions in MM2, for a total of twelve locations observed in this star forming region. 

These observations were done with a full-width half-power beam size of $20\arcsec$, %angular resolution of $0.26\arcsec$, 
spectral resolution of 1.1 km s$^{-1}$, and rms per channel of 2.0 mJy beam$^{-1}$. More details can be found in \citet{El_Abd_2019}, \citet{Hunter_2017}, \citet{McGuire2017} and \citet{Brogan_2018}.

\subsubsection{Matching parameter}
\label{sec:match}

For each comparison of a model with a set of observational data, a matching parameter, $m$, is calculated with the generalized form:
\begin{equation}
    m=\sqrt{m_1^2 + m_2^2 + ...}
	\label{eq:match1}
\end{equation}
\noindent where the number of terms depends on the number of molecular species used in the comparison. Each $m_i$ represents the quality of match for an individual quantity (molecule or molecular ratio) of index $i$, given by:
\begin{equation}
    m_i = \log\left(\frac{R_{\text{mod,i}}}{R_{\text{obs,i}}}\right).
	\label{eq:match2}
\end{equation}
The value $R_\text{mod,i}$ indicates a model-produced quantity and $R_\mathrm{obs,i}$ is the corresponding observed value. Since observational molecular abundance information is obtained as a set of column densities, while the chemical models instead produce local fractional abundances, $R_\text{obs,i}$ and $R_\mathrm{mod,i}$ are taken as molecular ratios with respect to some standard molecule. The value $m_i$ thus indicates the number of orders of magnitude by which the modeled molecular ratio deviates from the observed ratio, while $m$ is the root-mean-square of all such individual values, with $m=0$ being a perfect match. Other forms of matching parameter are of course possible; however, the chosen method ensures that, in the case of a non-perfect match, a model producing a single, large divergence is disfavored versus a model that produces many small divergences from the observed values.

Following G22 and related studies in which output data from the MAGICKAL model has been compared with gas-phase observational abundance values for hot cores/corinos, we use the {\it peak} gas-phase abundance produced by the stage-2 models for comparison with observations. Thus, for example, the modeled MF:GA ratio is calculated as the ratio of the peak gas-phase MF abundance to the peak gas-phase GA abundance. Although there is a degree of variation, most COMs reach their peak gas-phase abundances within a similar range of times/temperatures in the models. We therefore assume that this region/time of peak molecular abundance is representative of the observed emission from these molecules \citep[see e.g.][]{Belloche_2019}. Furthermore, the destruction of gas-phase COMs is dominated by ion-molecule reactions, and the rates of destruction for most COMs of interest in the models (including the C$_2$H$_4$O$_2$ isomers) are similar; this means that the post-peak molecular ratios remain fairly stable over time. In the absence of a spatial model of the distribution of the COMs, we therefore assume that the ratio of peak abundances between different molecules should be directly comparable with observed column density ratios.

The matching parameter described by Eqs.~(\ref{eq:match1}) and (\ref{eq:match2}) is used in several different ways to determine the quality of match of various models with particular observational datasets, as described below.

%%%%%%%%%%%%%%%%%%%%%%%
\begin{deluxetable*}{r l c c c c c l l c c c c}
\tabletypesize{\footnotesize}
\tablecaption{MF:GA and MF:AA ratios for the 12 sources in NGC 6334I observed by \citet{El_Abd_2019}. Listed alongside each source is an identification  number used exclusively in this paper to more easily indicate the best-match model for each source as shown in Figs.~\ref{fig:MFGA}--\ref{fig:MFAA}. Beside the observed molecular ratios are shown the best-match modeled ratios based on matching procedures \#1, \#2 and \#3, and the relevant model parameters. Note that procedure \#3 has the same best-match results as fitting procedure \#2 for the MM1 sources.\label{table2}}
\tablewidth{0pt}
\tablehead{\colhead{ID \#} & Source name & \multicolumn{2}{c}{Observed}      && \multicolumn{3}{c}{Best-Match Model}                    && \multicolumn{4}{c}{Model Conditions} \\
           \cline{3-4} \cline{6-8} \cline{10-13} \\
        &  & \colhead{MF:GA} & \colhead{MF:AA} && \colhead{MF:GA} & \colhead{MF:AA} & \colhead{$m$-value} && \colhead{A$_\text{v,init}$} & \colhead{\phantom{i0}$\text{n}_\text{H}$} & \colhead{$\zeta$} & \colhead{$t_{\text{wu}}$} \\
    & &  &&  &  &&&& \colhead{(mag.)} & \colhead{(cm$^{-3}$)} & \colhead{(s$^{-1}$)} & \colhead{(yr)} }
\startdata
\multicolumn{10}{l}{Matching Procedure \#1 - No Constraints}\\
    \hline
 1 & NGC 6334I MM1-i&    15.0&  21.8&& 15.0& 21.8& 0.002368 && 2 & $2\times10^{10}$ & $1.30\times10^{-15}$ & $2.00\times10^{4}$  \\
 2 & NGC 6334I MM1-ii&   15.6&  33.7&& 16.0& 35.9& 0.02372  && 3 & $2\times10^{10}$ & $1.30\times10^{-17}$ & $3.56\times10^{5}$  \\
 3 & NGC 6334I MM1-iii&  16.5&  9.80&& 15.2& 7.1 & 0.1459   && 3 & $2\times10^{10}$ & $4.11\times10^{-18}$ & $6.32\times10^{5}$  \\
 4 & NGC 6334I MM1-iv&   37.8&  87.5&& 39.2& 89.0& 0.01701  && 3 & $2\times10^{10}$ & $4.11\times10^{-17}$ & $3.56\times10^{5}$  \\
 5 & NGC 6334I MM1-v&    45.4&  6.17&& 63.9& 7.0 & 0.1573   && 3 & $2\times10^{10}$ & $4.11\times10^{-18}$ & $2.00\times10^{6}$  \\
 6 & NGC 6334I MM1-vi&   13.3&  12.7&& 11.2& 14.8& 0.1005   && 2 & $2\times10^{8} $ & $1.30\times10^{-15}$ & $1.12\times10^{5}$  \\
 7 & NGC 6334I MM1-vii&  25.2&  18.4&& 25.0& 17.6& 0.02000  && 3 & $2\times10^{10}$ & $1.30\times10^{-15}$ & $3.56\times10^{4}$  \\
 8 & NGC 6334I MM1-viii& 30.1&  34.3&& 33.1& 34.3& 0.04116  && 3 & $2\times10^{10}$ & $1.30\times10^{-17}$ & $6.43\times10^{5}$  \\
 9 & NGC 6334I MM1-ix&   35.3&  54.1&& 37.4& 51.8& 0.03132  && 2 & $2\times10^{6} $ & $1.30\times10^{-16}$ & $1.12\times10^{6}$  \\
10 & NGC 6334I MM2-i&   $>$158& 21.7&& 140&  18.2& 0.09528  && 2 & $2\times10^{10}$ & $1.30\times10^{-15}$ & $1.12\times10^{5}$  \\
11 & NGC 6334I MM2-ii&  $>$146& 15.8&& 235&  16.5& 0.01753  && 2 & $2\times10^{10}$ & $4.11\times10^{-18}$ & $2.00\times10^{6}$  \\
12 & NGC 6334I MM2-iii& $>$374& 28.3&& 577&  43.1& 0.1822   && 2 & $2\times10^{9} $ & $4.11\times10^{-16}$ & $3.56\times10^{5}$  \\
    \hline
\multicolumn{10}{l}{Matching Procedure \#2 - Fixed combination of $\zeta$ and $A_\mathrm{V,initial}$}\\
    \hline
 1 & NGC 6334I MM1-i&    15.0&  21.8&& 15.0& 21.8& 0.002368 && 2 & $2\times10^{10}$ & $1.30\times10^{-15}$ & $2.00\times10^{4}$  \\
 2 & NGC 6334I MM1-ii&   15.6&  33.7&& 15.0& 21.8& 0.2378   && 2 & $2\times10^{10}$ & $1.30\times10^{-15}$ & $2.00\times10^{4}$  \\
 3 & NGC 6334I MM1-iii&  16.5&  9.80&& 11.2& 14.8& 0.2461   && 2 & $2\times10^{8} $ & $1.30\times10^{-15}$ & $1.12\times10^{5}$  \\
 4 & NGC 6334I MM1-iv&   37.8&  87.5&& 32.2& 21.9& 0.6064   && 2 & $2\times10^{10}$ & $1.30\times10^{-15}$ & $3.56\times10^{4}$  \\
 5 & NGC 6334I MM1-v&    45.4&  6.17&& 38.1& 15.4& 0.4041   && 2 & $2\times10^{9} $ & $1.30\times10^{-15}$ & $1.12\times10^{5}$  \\
 6 & NGC 6334I MM1-vi&   13.3&  12.7&& 11.2& 14.8& 0.1005   && 2 & $2\times10^{8} $ & $1.30\times10^{-15}$ & $1.12\times10^{5}$  \\
 7 & NGC 6334I MM1-vii&  25.2&  18.4&& 19.0& 19.5& 0.1263   && 2 & $2\times10^{9} $ & $1.30\times10^{-15}$ & $6.32\times10^{4}$  \\
 8 & NGC 6334I MM1-viii& 30.1&  34.3&& 32.2& 21.9& 0.1974   && 2 & $2\times10^{10}$ & $1.30\times10^{-15}$ & $3.56\times10^{4}$  \\
 9 & NGC 6334I MM1-ix&   35.3&  54.1&& 32.2& 21.9& 0.3951   && 2 & $2\times10^{10}$ & $1.30\times10^{-15}$ & $3.56\times10^{4}$  \\
10 & NGC 6334I MM2-i&   $>$158& 21.7&& 140&  18.2& 0.09528  && 2 & $2\times10^{10}$ & $1.30\times10^{-15}$ & $1.12\times10^{5}$  \\
11 & NGC 6334I MM2-ii&  $>$146& 15.8&& 140&  18.2& 0.06224  && 2 & $2\times10^{10}$ & $1.30\times10^{-15}$ & $1.12\times10^{5}$  \\
12 & NGC 6334I MM2-iii& $>$374& 28.3&& 140&  18.2& 0.4690   && 2 & $2\times10^{10}$ & $1.30\times10^{-15}$ & $1.12\times10^{5}$  \\
    \hline
\multicolumn{10}{l}{Matching Procedure \#3 - Separate combination of $\zeta$ and $A_\mathrm{V,initial}$ combinations for MM1 and MM2} \\
    \hline
   & \multicolumn{10}{l}{MM1 sources: same as matching procedure \#2} \\
10 & NGC 6334I MM2-i&   $>$158& 21.7&& 235&  16.5& 0.1203   && 2 & $2\times10^{10}$ & $4.11\times10^{-18}$ & $2.00\times10^{6}$  \\
11 & NGC 6334I MM2-ii&  $>$146& 15.8&& 235&  16.5& 0.01753  && 2 & $2\times10^{10}$ & $4.11\times10^{-18}$ & $2.00\times10^{6}$  \\
12 & NGC 6334I MM2-iii& $>$374& 28.3&& 235&  16.5& 0.3102   && 2 & $2\times10^{10}$ & $4.11\times10^{-18}$ & $2.00\times10^{6}$  \\
\enddata
\end{deluxetable*}
%%%%%%%%%%%%%%%%%%%%%%%%%%%%%%%%%%%%%%%%%%%%%%%%%

\subsubsection{Comparisons with NGC 6334I}
\label{sec:ngc}

As the main focus of this work, comparisons of the chemical models with lines of sight toward NGC 6334I focus specifically on the MF:GA and MF:AA ratios. Thus, Eq.~(\ref{eq:match1}) requires only $m_1$ and $m_2$, which take values:
\begin{eqnarray}
    m_1 = \log\left(\frac{R_{\text{mod,MFGA}}}{R_{\text{obs,MFGA}}}\right) \nonumber \\ 
    m_2 = \log\left(\frac{R_{\text{mod,MFAA}}}{R_{\text{obs,MFAA}}}\right), \nonumber \label{eq:match-MFGA}
\end{eqnarray}
\noindent where $R_{\text{mod,MFGA}}$ is the MF:GA ratio from the model, and $R_{\text{obs,MFGA}}$ is the observed ratio, etc. In cases where the observational value of MF:GA is based on an upper limit for the column density of GA, models that achieve a ratio that matches {\it or exceeds} the observed value are assigned a perfect match parameter, i.e.~$m_{1} = 0$.

Table~\ref{table2} lists the twelve sight-lines observed in NGC 6334I and their respective column density ratios of MF to GA and MF to AA. For each of the sight-lines, a best-match model is sought. Three different procedures are used to determine these best matches, based on several further constraints, as summarized below:

\begin{enumerate}\label{}
    \item[1.]{No constraints: Matching procedure runs through all models and independently selects the best matching (lowest $m$) model for each sight-line.}
    \item[2.]{Uniform cosmic-ray ionization rate and initial extinction: Assumes that all sight-lines in NGC 6334I must have a shared $\zeta$-value. A shared $A_\mathrm{V,initial}$ value is also selected, to indicate a common background/ambient visual extinction for the cloud in its initial state.}
    \item[3.]{Separate MM1 and MM2: Same constraints as procedure 2, but allows different CRIR values for MM1 versus MM2 sight-lines.}
\end{enumerate}

For matching procedure \#1, all 756 stage-2 models are considered in one run through. Each model is analyzed for its peak abundance ratios of MF, GA, and AA and for each of the twelve sources; whichever model out of the 756 has the best MF:GA and MF:AA match compared to observations is deemed the best-match model. Although this method technically finds the best matches between models and observations, it ignores that the local CRIR in NGC 6334I may be the same or very similar between each sight-line.

For matching procedure \#2, the determination of the best-match model for each sight line is done separately for each $\zeta$-value tested in the models, in combination with one or other $A_\mathrm{V,initial}$ value, on the assumption that the same $\zeta$ applies throughout NGC 6334I, and all material originated from a cloud of the same or similar degree of exposure to external ionizing photons. For each of the 14 [$\zeta$,$A_\mathrm{V,initial}$] combinations, a model dataset of six density values and nine warm-up timescales (a total of 54 models) is compared with each of the twelve NGC 6334I sight-lines. The best match for each source is found within that set. An overall $m$-value is then determined for this [$\zeta$,$A_\mathrm{V,initial}$] combination, whereby the $m$-values of the best matching models for each of the sources are combined according to Eq.~\ref{eq:match1}. The [$\zeta$,$A_\mathrm{V,initial}$] combination with the lowest overall $m$-value is considered the best overall match to the observations. The best-match model for a particular source is then considered to be the one drawn from this overall best-match combination.

For matching procedure \#3, the same constraints as matching procedure \#2 are used, but sources MM1 and MM2 are considered separately, allowing them to settle on independent best-match [$\zeta$,$A_\mathrm{V,initial}$] combinations.

The above-described matching procedures are based on comparisons of molecular ratios, but by themselves they do not take account of the absolute abundance values of individual species. In principle, this could lead to models being chosen as a good/best match while not otherwise being suitable as a description of hot-core chemistry. To avoid the consideration of models with unrealistically low molecular abundances, thresholds were set for the two key molecules methyl formate and methanol. 
Models for which the gas-phase methyl formate abundance falls below $10^{-9}\text{n}_\text{H}$, are removed from consideration as best-match models. Methyl formate abundances in hot cores are typically on the order of 10$^{-8}\text{n}_\text{H}$ or higher \citep{bisschop_2007}. As a further constraint, we also remove models with peak methanol abundances less than $10^{-6}\text{n}_\text{H}$, although models that fail one condition typically fail both.

%%%%%%%%%%%%%%%%%%%%%%%
\begin{figure}
\centering
	\includegraphics[width=\columnwidth]{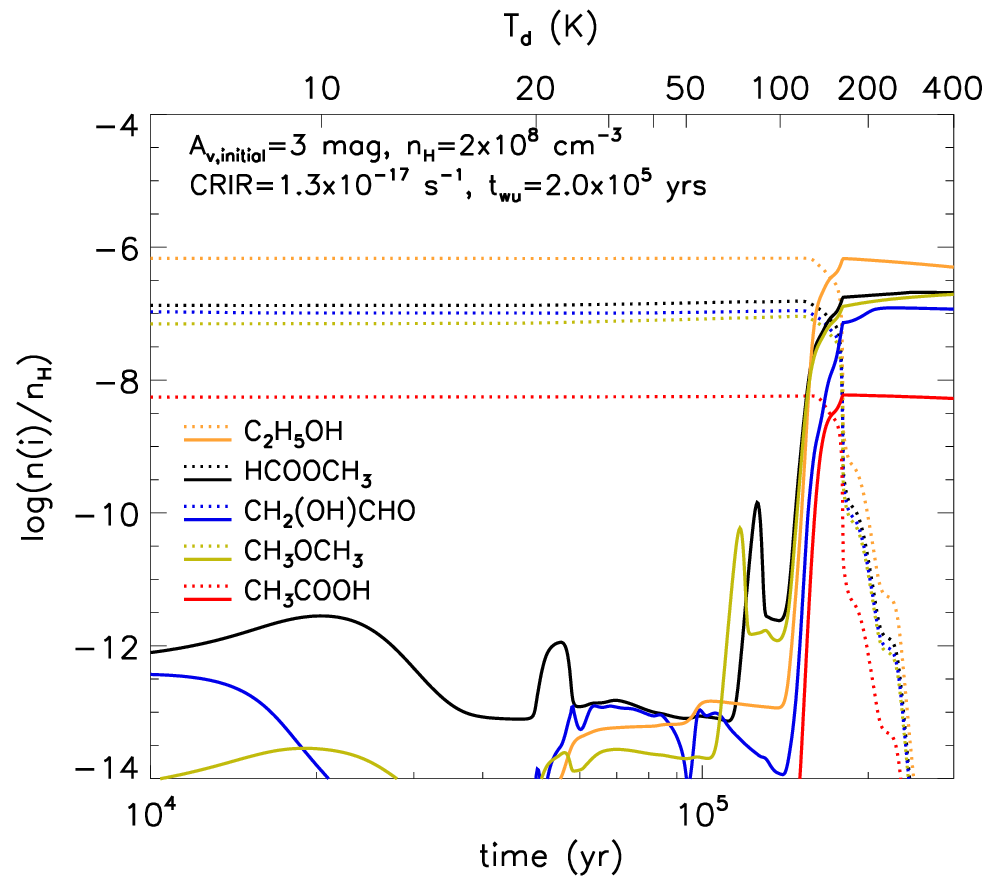}
    \caption{Selected fractional abundances produced by the model with standard physical conditions as considered by \citet{Garrod_2022}} (shown in Table~\ref{table:conds} where A$_\text{v,init}=3$ mag, n$_\text{H,final} = 2\times10^{8}$ cm$^{-3}$, CRIR = $1.3\times10^{-17}$ s$^{-1}$, and t$_\text{wu}=2\times10^{5}$ yr). Gas (solid lines) and grain (dotted lines) abundances of MF (black), GA (blue), AA (red), ethanol (orange), and dimethyl ether (green) with respect to total hydrogen are shown.\label{fig:abundances1}
\end{figure}
%%%%%%%%%%%%%%%%%%%%%%%

%%%%%%%%%%%%%%%%%%%%%%%%%%%%%%%%%%%%%%%%
%%% RESULTS %%%%%%%%%%%%%%%%%%%%%%%%%%%%
%%%%%%%%%%%%%%%%%%%%%%%%%%%%%%%%%%%%%%%%
\section{Results}
\label{general}

\begin{figure}
	\includegraphics[width=\columnwidth]{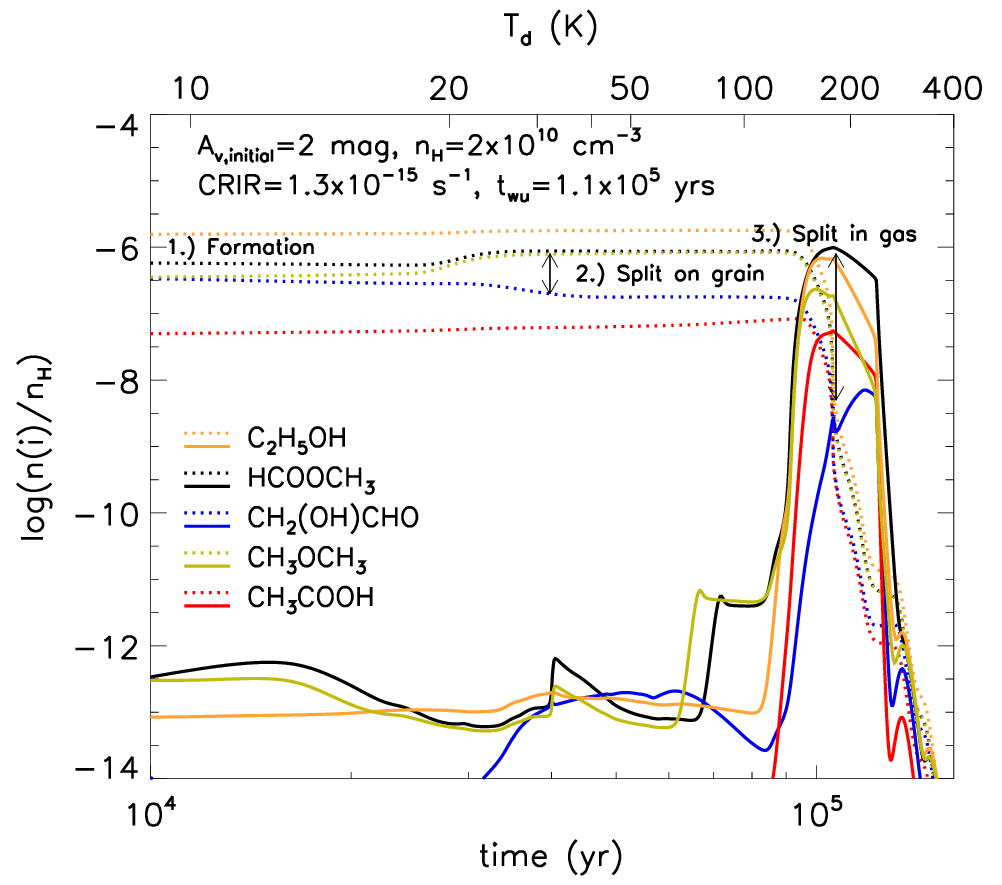}
    \caption{As Fig.~\ref{fig:abundances1}, with results from an alternative model with higher gas density ($2\times10^{10}$ cm$^{-3}$), CRIR ($1.3\times10^{-15}$ s$^{-1}$), lower initial visual extinction (2 mag), and shorter warm-up timescale ($1.1\times10^{5}$ yrs), producing a more substantial divergence in grain-surface and gas-phase MF and GA abundances. The model shown is the best-match model for NGC 6334I MM2 sources, using matching procedure \#2. Color scheme as per Fig.~\ref{fig:abundances1}. Note the different conditions from Fig.~\ref{fig:abundances1} as stated at the top of the figure.\label{fig:abundances2}}
\end{figure}

\begin{figure}
	\includegraphics[width=\columnwidth]{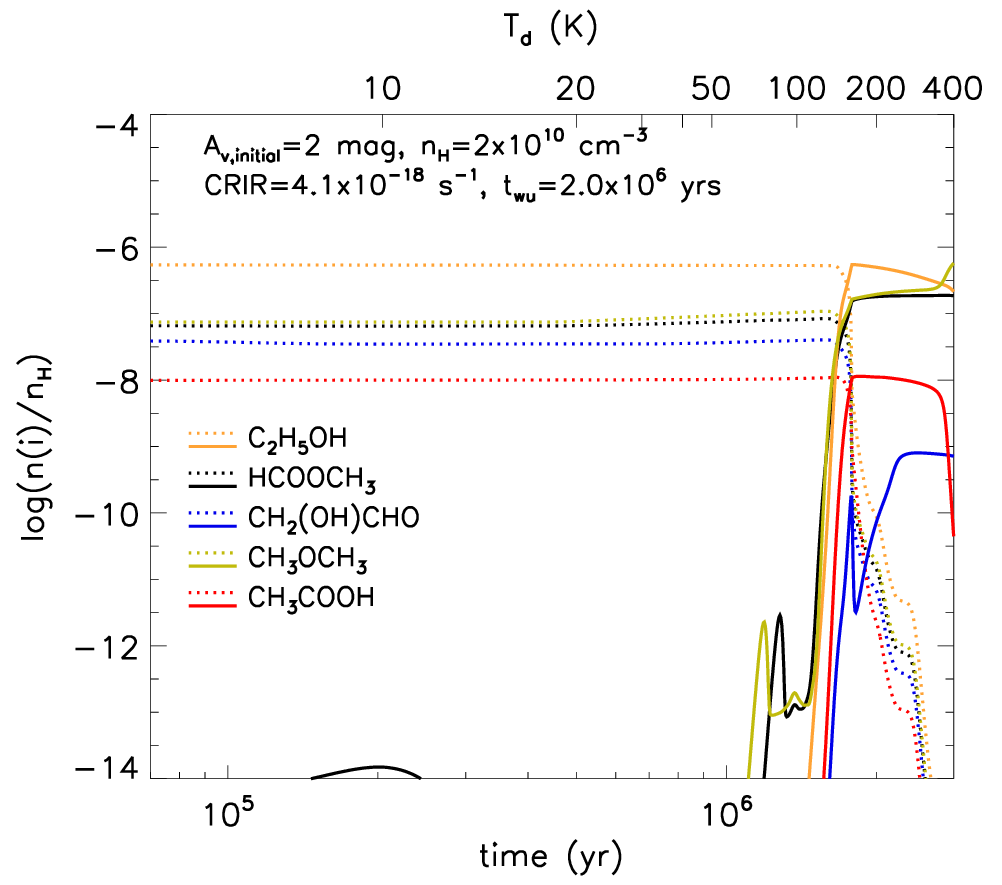}
    \caption{As Fig.~\ref{fig:abundances1}, with results from an alternative model that is the best match for NGC 6334I MM2 sources with matching procedure \#3. Same color scheme as per Fig.~\ref{fig:abundances1} and Fig.~\ref{fig:abundances2}. Note the different conditions from Fig.~\ref{fig:abundances2} as stated at the top of the figure, with a lower CRIR ($4.1\times10^{-18}$ s$^{-1}$) and longer warm-up timescale ($2.0\times10^{6}$ yrs). The gas density remains high at $2\times10^{10}$ cm$^{-3}$ and the initial visual extinction at 2 mag.} \label{fig:abundances3}
\end{figure}

In order to gain insight into the chemistry influencing MF (HCOOCH$_3$), GA (CH$_2$(OH)CHO), and AA (CH$_3$COOH) in the models, we first consider the time-dependent abundances of these isomers, as well as the chemically related COMs ethanol (C$_2$H$_5$OH) and dimethyl ether (CH$_3$OCH$_3$). We focus here on three particular models from the grid, corresponding to (i) the ``standard'' conditions used by G22 (and shown in Table~\ref{table:conds}); (ii) a model with different conditions including more extreme density and CRIR values; and (iii) a model with somewhat lower CRIR, high gas density, and a long warm-up timescale. The first of the three models (Fig.~\ref{fig:abundances1}) is essentially the same as that presented by G22, for which a broader range of chemical abundance plots may be found in their Figures 8 and 9 (stage 1) and their Figure 11 (stage 2). The question of which model best matches each observed source is explored below in Sec.~\ref{sec:matching}; however, we note that model (ii) considered here (Fig.~\ref{fig:abundances2}) corresponds to the best match for the NGC 6334I MM2 sources using matching procedure \#2. Model (iii) (Fig.~\ref{fig:abundances3}) is the best match for MM2 sources using matching procedure \#3 (see Sec.~\ref{sec:matching} for further discussion). Here we describe the broad features and differences in the abundance behavior shown in the three figures, digging deeper into the underlying chemistry in Secs.~\ref{form-mech}--\ref{split2}.

%%%%%%%%%%%%%%%%%%%%%%%
%%%%%%%%%%%%%%%%%%%%%%%

%%%%%%%%%%%%%%%%%%%%%%%
%%%%%%%%%%%%%%%%%%%%%%%

Fig.~\ref{fig:abundances1} shows COM abundances during the stage-2 (warm-up) evolution of the ``standard'' model (i) setup. For consistency between models, results are plotted beginning at time $10^4$~yr, when the dust and gas temperatures are $\sim$8~K and 10~K, respectively. The plot continues until the end-time of the model, which occurs at $\sim$$2.85 \times 10^5$ yr, when gas and dust temperatures of 400~K are reached.

The COMs shown in the plot are largely formed on the dust grains during stage 1 (a depiction of the net production rates through stages 1 and 2 may be seen in Figure 13 of G22). The early production of each of the grain-surface ice species (dotted lines) derives a substantial contribution from radical-association reactions on the surfaces as the ices build up under low-temperature and high gas-density conditions. In the case of AA and ethanol, photochemistry in the young, thin, grain-surface ices is also important. Solid-phase GA abundances are intertwined with the abundances of the related glyoxal (HCOCHO) and ethylene glycol (HOCH$_2$CH$_2$OH) molecules, via H-atom addition and abstraction. The solid-phase abundances of these COMs are stable until the ices begin to sublimate.

The gas-phase abundances of the COMs (solid lines of the same colors) reach their peaks at temperatures above $\sim$120~K, when the ices begin to sublimate. Some of the gas-phase abundances do not reach their full peak values until the large water-ice component has fully desorbed, which occurs at a temperature around 160~K. Release of some species at yet higher temperatures can also occur, either due to their own high binding energies, or because of trapping beneath species that have not yet desorbed. Gradual declines in gas-phase abundances toward the end of the model are caused by protonation of the COMs by abundant molecular ions such as \ce{H3O+}; protonated COMs are destroyed by dissociative recombination with electrons. The peak gas-phase abundances of the \ce{C2H4O2} isomers in Fig.~\ref{fig:abundances1} are in the correct order with respect to observational values, i.e. MF is most abundant and AA is the least abundant. The MF:GA ratio is $\sim$1.7.

%%%%%%%%%%%%%%%%%%%%%%%
\begin{figure*}
\centering
	\includegraphics[width=0.8\textwidth]{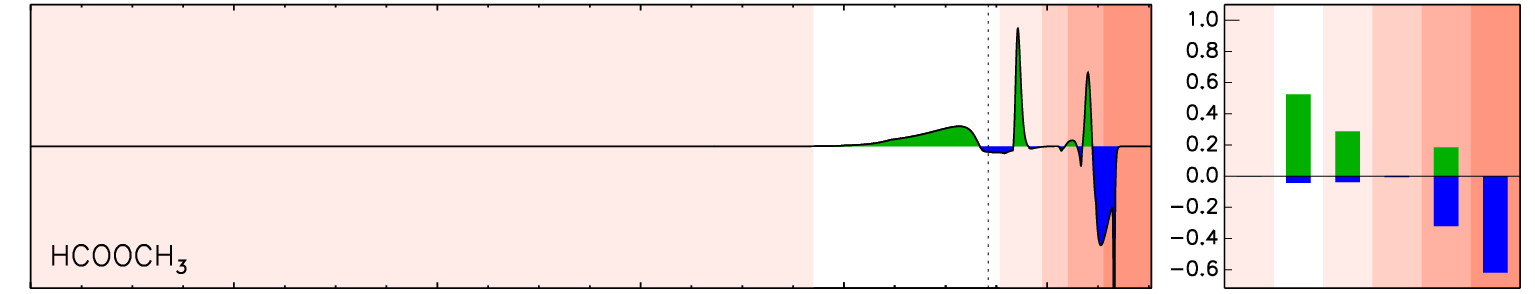}
	\linebreak
	\includegraphics[width=0.8\textwidth]{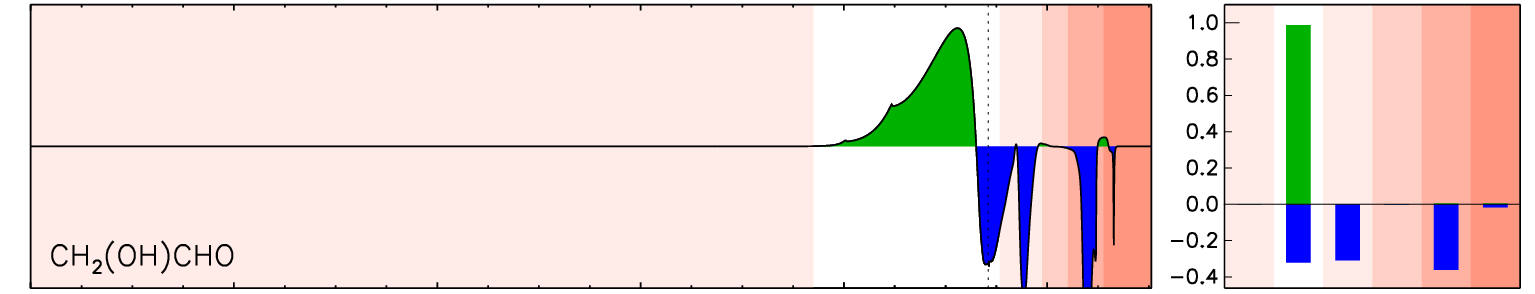}
	\linebreak
	\includegraphics[width=0.8\textwidth]{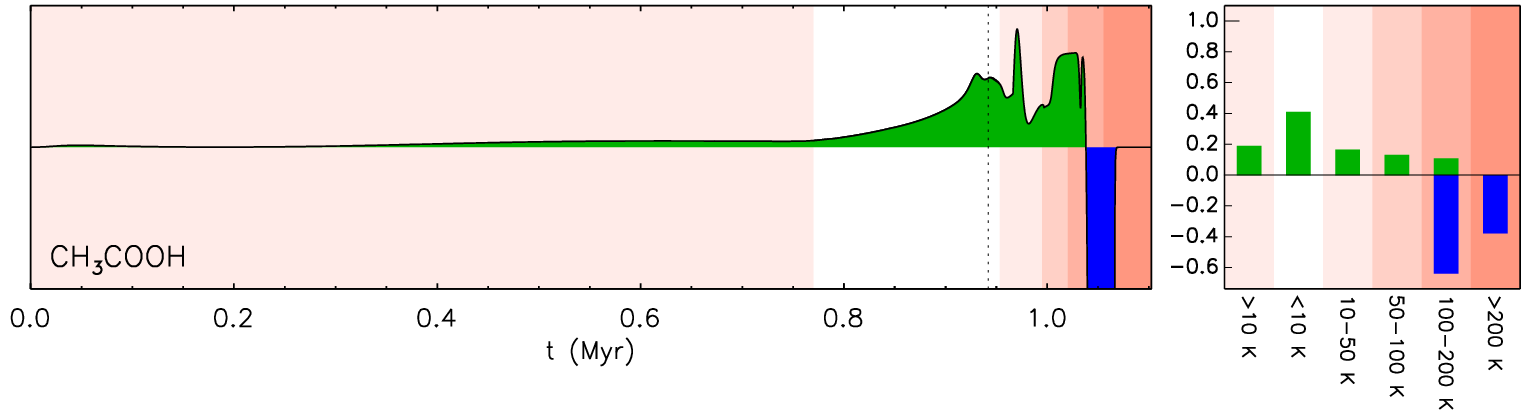}
    \caption{{\bf Left panels:} Net rate of change of the three C$_2$H$_4$O$_2$ isomers, as a function of time, for the best-matching model to the NGC 6334I MM2 abundances using match parameter \#2. Rate of change for each molecule is summed over all phases, and is shown in arbitrary units. Net formation is highlighted in green and net destruction in blue. The dotted vertical line indicates the transition from stage 1 to stage 2. {\bf Right panels:} The same data is shown as a fraction of the total net formation (i.e. the sum of all green areas), for each of six temperature regimes occurring through stages 1 and 2. Different background shading indicates each temperature regime. Destruction bars (blue) are also normalized to the total net formation.}
    \label{fig:productionrates}
\end{figure*}
%%%%%%%%%%%%%%%%%%%%%%%

Fig.~\ref{fig:abundances2}, showing results from model (ii) -- which has 100 times higher stage-2 gas density and 100 times higher cosmic ray ionization rate -- reveals some interesting differences. Firstly, all of the solid-phase COM abundances have been enhanced during their earlier formation in stage 1; the stronger secondary UV field, produced by CR excitation of gas-phase H$_2$, leads to a greater abundance of radicals on the grain surfaces and in the ices, which then recombine to form more COMs. Meanwhile, toward the end of stage 2, following the desorption of the COMs into the gas phase, the higher CRIR drives a more rapid ion-molecule chemistry. Protonation and dissociative recombination destroy these species completely before the end of the model is reached, in spite of the timescale of this model being slightly shorter than the ``standard'' value of model (i).

Two other important features are also notable; whereas in model (i) the solid-phase abundances of MF and GA seem to track in parallel with each other, more or less preserving their abundances from stage 1, in model (ii) these two species diverge at a temperature around 20--30~K, with MF increasing in abundance while GA decreases. Dimethyl ether behaves similarly to MF at this point in the model.

An even greater divergence is then seen when the ices desorb into the gas phase; while MF abundance is essentially preserved between its peak solid-phase abundance prior to desorption and its subsequent gas-phase peak value, GA drops drastically during the desorption period, so that it never reaches the same abundance in the gas phase as it had in the ice just prior to desorption. The gas-phase MF:GA ratio in this model is 140:1.

In Fig.~\ref{fig:abundances3}, in which the results of the high-density, long warm-up timescale model (iii) are shown, the solid-phase abundances are again more stable until sublimation occurs, although the abundances are generally lower (excluding dimethyl ether) than in the ``standard'' model (i), due to the slightly lower CRIR that results in weaker stage-1 production of radicals. The large split in peak gas-phase abundances of MF and GA is retained in this high-density model, with a MF:GA ratio of 235:1. Methyl formate abundance actually grows during the period of strong desorption, rising above the prior solid-phase abundance. Although the CRIR is a little lower than standard, the long timescale of this model means that ion-molecule chemistry diminishes some of the molecules more substantially by the end of the model run than in the ``standard'' case.

\subsection{Early Formation Mechanisms}
\label{form-mech}

Here we provide a more detailed description of the formation mechanisms of the \ce{C2H4O2} isomers occurring mainly during stage 1 (collapse), which are in general responsible for much of the formation of these molecules under typical conditions. 

Figure \ref{fig:productionrates} (left panels) shows the net rate of formation (green) and destruction (blue) for MF, GA, and AA at times throughout model (ii), from the beginning of stage 1 through the end of stage 2. These rates correspond to the sum of formation and destruction of the same molecule over all phases (i.e. gas, grain-surface and bulk-ice). The dotted line in the plot separates stage 1 from stage 2. The right panels of the same figure indicate the integrated formation and destruction in several broad temperature regimes, normalized to the total formation. The first two regimes correspond mainly to stage 1 of the models; most of the net production (shown in green) of MF, GA, and AA is seen to occur during stage 1, when dust temperatures are below 10 K and the gas density is reaching its peak, resulting in rapid adsorption of gas-phase atoms and molecules onto the grain surfaces.

At these early stages in all three models, methyl formate is predominantly formed on the grain surfaces through the 3-B reaction of radicals HCO and CH$_{3}$O, in which either the HCO or the CH$_3$O radical is formed close to its reaction partner. For example, 
\begin{subequations}
\begin{eqnarray}
\ce{[H + CO] + CH_{3}O \rightarrow HCO + CH_{3}O}  \\
\ce{HCO + CH_{3}O \rightarrow HCOOCH_{3}}
\end{eqnarray}
\end{subequations}
In this case, the initiating reaction between H and CO occurs through a typical diffusive mechanism, following the adsorption of H onto the grain surface from the gas phase.

In model (ii), the high CRIR results in a large gas-phase abundance of atomic H (due to CR-induced dissociation of H$_2$), around 100 times the model (i) value. The rate of H adsorption onto the grains is commensurately higher than in model (i), leading to more efficient hydrogenation of CO and H$_2$CO, and a higher production rate for MF. The final gas density in model (ii) is also higher by a factor of 100, but this high density is reached very late in stage 1, after much of the gas-phase CO used to form COMs has already frozen out onto the grains.

As discussed by G22, the chemical network includes the 3-BEF mechanism, which in this case allows the energy of formation of the just-formed CH$_3$O radical to overcome the activation energy barrier to immediate reaction with nearby CO. The product of this reaction, the radical CH$_3$OCO, may then recombine with atomic H through a normal diffusive reaction to produce MF:
\begin{equation}
\label{reaction4}
\ce{H + CH$_3$OCO$\rightarrow$ HCOOCH$_3$}
\end{equation}
Under certain circumstances, this process may temporarily become the dominant MF-formation process in the model, but overall it contributes only on the order of a few percent maximum to the total production of methyl formate.

As seen in Fig.~\ref{fig:productionrates}, in model (ii) there is some substantial production of MF at later times and higher temperatures during stage 2. This is described in detail in Secs.~\ref{split1} and \ref{split2}.

For glycolaldehyde, the formation mechanisms are more complicated, as there is interconversion between glyoxal (HCOCHO), glycolaldehyde (GA), and ethylene glycol ((CH$_2$OH)$_2$), which affects the overall production and destruction of GA. Atomic H may react with each of the three species, mediated by activation energy barriers, to add or abstract a hydrogen atom, producing a radical. Reaction of H with such a radical may further add or subtract an H atom to the structure (without an activation barrier). As discussed by G22, the degree of interconversion is thus dependent on a number of factors, including how rapidly the ice is building up on the grain surface. However, ultimately it is the radical recombination reactions:
\begin{eqnarray}
\ce{HCO + HCO \rightarrow HCOCHO} \\
\ce{HCO + CH2OH \rightarrow CH2(OH)CHO} \\
\ce{CH2OH + CH2OH \rightarrow (CH2OH)2}
\end{eqnarray}
that add new material into this triad of related species. During the cold stage 1 when most of this material is formed, each of these reactions occurs mainly via the 3-B mechanism. The CH$_2$OH is formed mostly through the barrier-mediated diffusive surface reaction
\begin{equation}
\ce{H + CH_{3}OH \rightarrow CH_{2}OH + H_{2}}
\end{equation}
More detail on the interconversion between glyoxal, glycolaldehyde, and ethylene glycol can be found in G22. Essentially all of the net positive production rates of GA in model (ii) occur at the end of the cold collapse stage; GA production continues at later times in moderate amounts, but it is not sufficient to overcome photoprocessing in the ice, as well other destructive mechanisms described below.

AA production occurs mainly during stage 1 also; however, unlike MF and GA, the earliest stages of ice formation at low visual extinction also provide an opportunity for external UV to drive AA production. Much of this production occurs through the two reactions
\begin{eqnarray}
\ce{CH3CO} + \ce{OH} \rightarrow \ce{CH3COOH} \\
\ce{CH3} + \ce{COOH} \rightarrow \ce{CH3COOH}
\end{eqnarray}
Several mechanisms can produce the radical CH$_3$CO, including H-abstraction from acetaldehyde (CH$_3$CHO) and the barrier-mediated reaction of CH$_3$ with CO, while OH abundance is maintained by the photodissociation of water ice. The CH$_3$ radical is an intermediate in the hydrogenation of atomic C to form methane, while the COOH radical can be produced by the destruction of formic acid (HCOOH).

The high CRIR in model (ii) has more dissociating photons than in the other two models, even at high visual extinction, meaning that PDI-driven AA production continues strongly even into stage 2. This behavior, shown visually in Fig.~\ref{fig:productionrates} may be contrasted with that presented by G22 (in their Figure~13) for ``standard'' conditions, in which most AA is formed at very early times in stage 1 when external UV is still able to penetrate.

\subsection{Grain-Surface Split in MF and GA Abundances}
\label{split1}

Early in the stage-2 evolution of model (ii), as seen in Fig.~\ref{fig:abundances2}, MF and GA abundances stay relatively steady in relation to each other, showing a slight downward trend with time that is due to gradual CR-induced photodissociation of the ices. Then, as marked in the figure, beginning at a temperature of $\sim$17~K (or a time in stage 2 around $2.5 \times 10^{4}$~yr), a split appears in the behavior of grain-surface GA and MF; GA abundances falls, while that of MF (and of dimethyl ether) rises. This divergence is complete by around 35~K ($\sim$$4.5 \times 10^{4}$~yr). The new MF:GA ratio is retained in the ice until strong desorption occurs at temperatures $>$100~K. Note that the grain-surface abundances shown in the figures correspond to the total ice present in the bulk ice and the outer surface layer combined.

This divergence in grain-surface abundances is a bulk-ice chemistry effect. Below $\sim$17~K, bulk diffusion of atomic H is slow, and is governed by tunneling between interstitial sites in the ice matrix. Above this temperature, thermal diffusion takes over, with a rate that grows rapidly as the grains become warmer. This change in H-diffusion behavior occurs in all of the stage-2 models; however, in model (ii), with its 100-times higher CRIR than the standard value, there is also a much greater production rate for atomic H, making the change in H mobility in the bulk ice much more important. This combination of high production-rate and high mobility of atomic H manifests in several chemical effects in model (ii).

We first consider the rise in MF abundance. There is more MF present at the beginning of stage 2 in model (ii) than there is in e.g. model (i), for reasons described above. But there is also a correspondingly higher abundance of the related radical \ce{COOCH3}. At low temperatures ($<$17~K) the latter species is formed in the bulk ice when MF is attacked by atomic H at the aldehyde end of the molecule. Although this reaction is mediated by an activation energy barrier \citep[$\sim$4000~K in this model;][]{Good2002}, the site-to-site diffusion rate of atomic H is low enough that reaction through the barrier is competitive with the alternative outcome in which the H atom simply diffuses away before reaction occurs. Thus, when an H-atom meets MF in the ice, it efficiently produces the \ce{CH3OCO} radical: 
\begin{equation}
\ce{H + HCOOCH$_3$ $\rightarrow$ CH$_3$OCO + H$_2$} \\
\end{equation}
When another H atom meets with this radical, MF is reformed through reaction \ref{reaction4}, which is assumed to have no substantial activation energy barrier.

The H in the bulk ice is formed exclusively through CR-induced photodissociation of the various molecules in the ice, dominated by H$_2$O. At low temperatures, the formation and destruction of \ce{HCOOCH3} and \ce{COOCH3} through reactions with H are fairly slow, and the ratio of MF to its related radical is stable.

As the temperature climbs above 17~K, thermal diffusion of atomic H takes hold; the efficiency of the H-abstraction reaction falls, because when an H-atom meets a MF molecule, diffusion away from MF becomes more probable than reaction. At the same time, the rate for the barrierless reaction of H with \ce{COOCH3} increases with temperature, due to faster diffusion.

In model (ii), in which there is a lot of H being formed through CR-induced photodissociation of various species, the balance shifts strongly at T$>$17~K, such that most \ce{COOCH3} in the ice is converted to MF without being converted back. Furthermore, in model (ii), there is already a lot of this radical present in the ice (\ce{HCOOCH3}:\ce{COOCH3}$\simeq$2) at the beginning of stage 1, thus the MF abundance rises substantially when this conversion takes place. In model (i) and other models with relatively low CRIR, there is less H being produced in the ice during both stages 1 and 2, meaning that the H-abstraction/addition cycle is already much slower (by $\sim$100 times). As a result, there is a lower initial ratio of \ce{HCOOCH3} to \ce{COOCH3} in the ice, while the change in diffusion behavior with temperature has only a very mild influence on that ratio during the stage-2 warm-up.

In Figure \ref{fig:productionrates}, which shows production rates from model (ii), we see net formation of MF not only in the <10K regime but also again in the 10-50K (and 100-200K) temperature regime, occurring sharply in the $\sim$17--30~K range.

A similar effect involving H addition and abstraction is also seen to influence the abundance of solid-phase dimethyl ether (\ce{CH3OCH3}) in model (ii).

For GA, much of the fall in its ice abundance in model (ii) occurs at around 25--35~K. In relative terms, this decline is a little less drastic than the rise in MF. In this case, conversion into ethylene glycol ((CH$_2$OH)$_2$) and, to a lesser extent, glyoxal (OHCCHO) accounts for much of the loss in glycoladehyde. The lowest barriers against either abstraction or addition of atomic H (several radical structures are possible) are quite low, on the order of 1000~K or less. This means that, upon the meeting of H and GA in the ice, a reaction is the most likely outcome versus diffusive escape of the H atom, even at temperatures as high as $\sim$35~K. Until this temperature approaches, conversion of GA continues effectively, while above this temperature, the diffusion of H away from GA is more likely than overcoming the barrier to addition or abstraction. While the H could in principle return to the GA to attempt reaction again, it is more likely to meet and react with a radical elsewhere in the ice, which would occur immediately upon meeting, rather than being hampered by a barrier.

In Figure \ref{fig:productionrates}, we see that for GA there is a net destruction at <10K and 10-50K, with the latter negative spike being representative of the destruction of GA described here.

For the models with low CRIR, such as model (i), again the lower production rate of atomic H in the bulk ice means that the conversion of GA is much slower than in model (ii), such that it has very little practical effect on its abundance.

Thus, the divergence in grain-surface abundances of MF and GA in models with high CRIR is driven by the changing mobility of H atoms in the ices, amplified by the greater production of H by the CR-induced photodissociation of other molecules. The divergence stabilizes again when the H atoms become too mobile for barrier-mediated reactions to be efficient versus recombination with radicals in the ice.

As described by G22, we note that in the present models diffusive radical-radical reactions play only a small part in the production of the COMs of interest here, although non-diffusive reactions between radicals are very influential. The diffusion of atomic H, either on the ice surface or in the bulk ice, is important in driving many of these non-diffusive reactions, by producing new radicals in close proximity to others.

\subsection{Gas-Phase Split in MF and GA Abundances}
\label{split2}

The split in gas-phase abundances of MF and GA in model (ii), also marked in Fig.~\ref{fig:abundances2}, is more drastic than the earlier grain-surface split. This divergence is ultimately related less to the higher CRIR of model (ii) than to the higher gas density. However, the reason that MF and GA are differently affected is their different binding energies.

As the dust temperature begins to rise during stage 2, some of the more volatile components of the ice start to desorb, such as CO and CH$_4$. However, in this three-phase chemical model, as material is lost from the surface monolayer, material in the bulk ice beneath is exposed, becoming the new surface. While some of this material is also composed of volatiles, other molecules with much higher binding energies, such as water, are also present. Thus, as volatiles leave, the upper layer becomes more concentrated in non-volatiles, leading to a trapping effect that acts to retain species of lower binding energy that would otherwise desorb at that temperature. The more volatile species are trapped until temperatures are achieved at which the less volatile species holding them in place may themselves thermally desorb.

Because of this effect, much of the total ice content is retained on the grain surfaces until temperatures around 120~K are achieved (see G22), when water itself begins to sublimate substantially from the grains. In this model, water has a binding energy of 5700~K, while the value for MF is 4210~K \citep{Burke15}, based on desorption from a water-ice surface. Hence, as soon as water in the upper monolayer desorbs, any uncovered MF immediately desorbs, giving it a very short lifetime on the surface. Although it takes some time for the water to desorb entirely from the grains (the process being essentially complete at a temperature of $\sim$160~K; G22), MF nevertheless desorbs as soon as it is exposed on the surface.

Conversely, GA has much greater binding energy \citep[5630~K;][]{Burke15} compared with MF, similar to that of water in this model, due to the hydroxyl group on the molecule that allows it to form hydrogen bonds with the water surface. As a result, GA has a lifetime on the water surface that is similar to water itself, taking some time to fully desorb. 

During its much longer residence time, GA comes under a sustained chemical attack before it can leave the surface. In particular, H atoms from the gas phase may stick briefly to the surface, where they can diffuse to meet reactants, including GA. In spite of the activation energy barriers, H may react with GA to abstract another H atom, forming H$_2$. Much of the abstraction occurs at the aldehyde end of the molecule, producing the CH$_2$(OH)CO radical. The latter may recombine with another H atom to re-form GA, or it may thermally desorb with a slightly lower binding energy than GA (as assumed in this model). Alternatively, an H-atom may be abstracted from the other end of the GA molecule, producing a radical that can react with another H atom to produce glyoxal (OHCCHO). The latter may also desorb more easily from the icy grain surface than GA.

Thus, much of the what was previously glycolaldehyde actually desorbs in a different form, so that the post-desorption gas-phase GA abundance is lower than the pre-desorption solid-phase abundance. Other radicals on the surface, freed by the gradual desorption of the water overlayers, may also react with the radicals formed from the hydrogen-atom attack on GA molecules, producing alternative species with higher binding energies. 

Because the attacking H atoms derive from the gas phase, the degree of destruction experienced by GA molecules prior to complete desorption from the grains is strongly dependent on the gas density, which determines the rate of adsorption. The high CRIR also increases the abundance of H atoms in the gas, which has a lesser but significant effect on the rate of GA destruction.

In summary, the trapping of MF beneath the water-ice surface, at temperatures above its own desorption temperature, means that the lifetime of any individual MF molecule on the surface is very short. Conversely, because GA is retained on the grains for longer, i.e. with individually longer surface lifetimes, it is strongly susceptible to attack by H atoms adsorbed from the gas, before it is able to escape. The effect of H-atom attack on GA is accelerated primarily by the higher gas density, and secondarily by the higher CRIR, in model (ii).

We note also that in model (iii), as shown in Fig.~\ref{fig:abundances3}, a similar gas-phase split is seen between MF and GA. In this case, although the CRIR is much smaller, the gas density is high, while the warm-up timescale is long. This leads to a longer period over which GA is exposed on the hot grains prior to desorption, while the high density supplies atomic H rapidly to the surface.

The MF abundance in model (iii) is seen to increase further during the desorption period, beyond what is even present on the grains. A similar -- and related -- effect is seen for dimethyl ether at around the same time. It should be noted that the abundances of both MF and DME in the ices in model (iii) are lower than those in model (ii). However, the very low CRIR in model (iii) means that the radicals CH$_3$OCO and CH$_3$OCH$_2$ that build up in the ice are far less likely to encounter an H atom with which they may recombine, allowing them to build up in the ices until late times. When each of these radicals becomes exposed to the surface, and to the accompanying flux of H atoms from the gas phase, they rapidly recombine to produce MF and DME, respectively, which then thermally desorb.

\subsection{General Effects of each Parameter on MF:GA:AA ratios}
\label{effects}

The effects of cosmic-ray ionization rate, initial visual extinction, final hydrogen number density and warm-up timescale on the broad trends in MF:GA and MF:AA ratios are now considered.

\begin{sidewaysfigure}
\centering
  \includegraphics[trim={0.3cm 0.1cm 0.07cm 0}, clip, height=3.6cm]{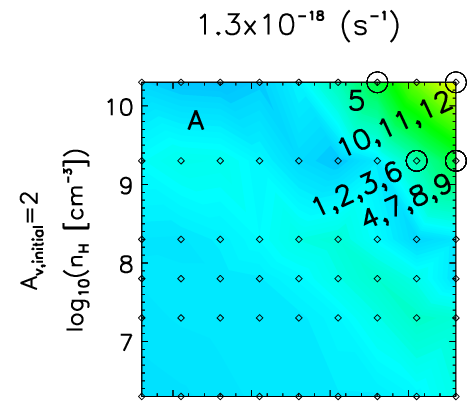}
  \includegraphics[trim={0.22cm 0.1cm 0.07cm 0}, clip, height=3.6cm]{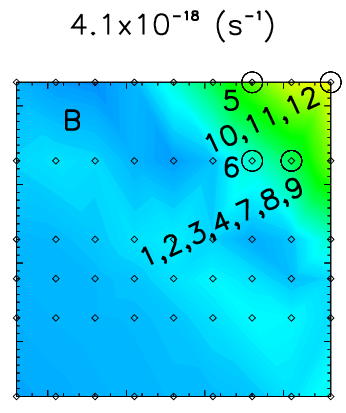}
  \includegraphics[trim={0.22cm 0.1cm 0.07cm 0}, clip, height=3.6cm]{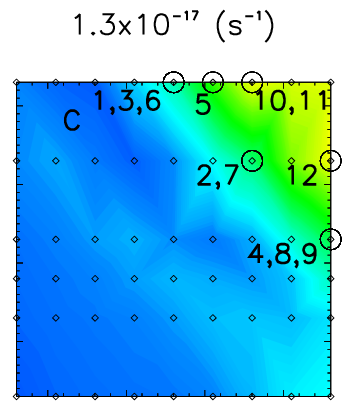}
  \includegraphics[trim={0.22cm 0.1cm 0.07cm 0}, clip, height=3.6cm]{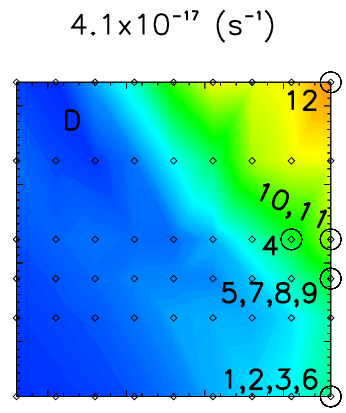}
  \includegraphics[trim={0.22cm 0.1cm 0.07cm 0}, clip, height=3.6cm]{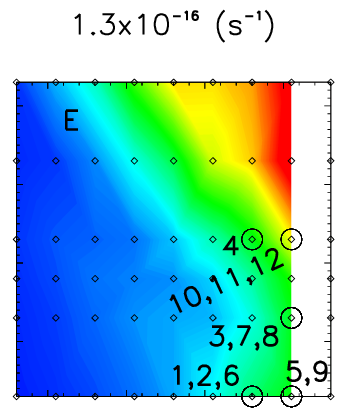}
  \includegraphics[trim={0.22cm 0.1cm 0.2cm 0}, clip, height=3.6cm]{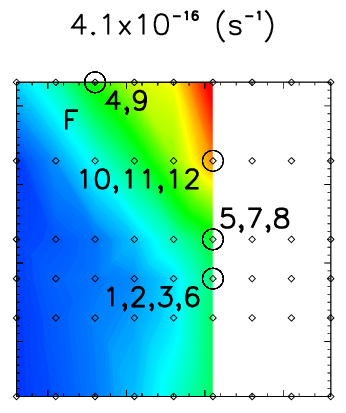}
  \includegraphics[trim={0.1cm 0.1cm 0.6cm 0}, clip, height=3.6cm]{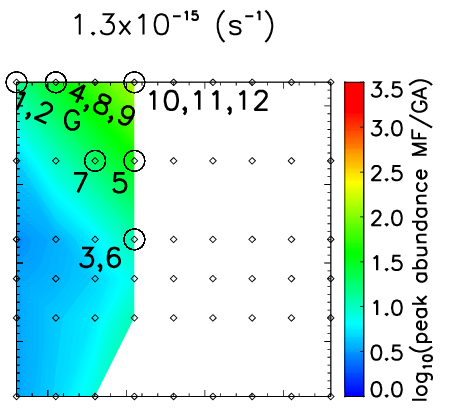}\\
  \includegraphics[trim={0.3cm 0.1cm 0.07cm 0}, clip, height=3.6cm]{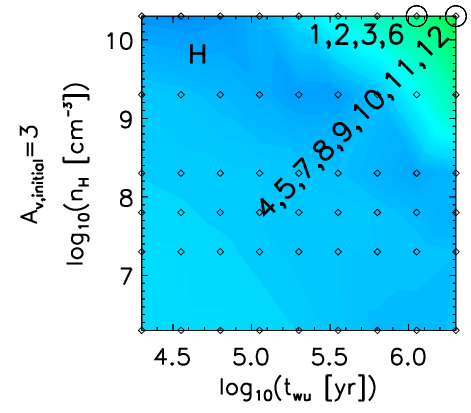}
  \includegraphics[trim={0.22cm 0.1cm 0.07cm 0}, clip, height=3.6cm]{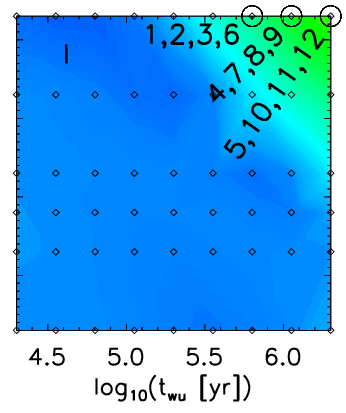}
  \includegraphics[trim={0.22cm 0.1cm 0.07cm 0}, clip, height=3.6cm]{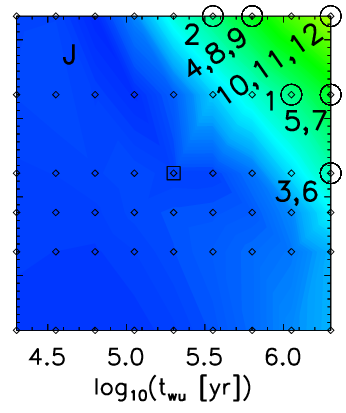}
  \includegraphics[trim={0.22cm 0.1cm 0.07cm 0}, clip, height=3.6cm]{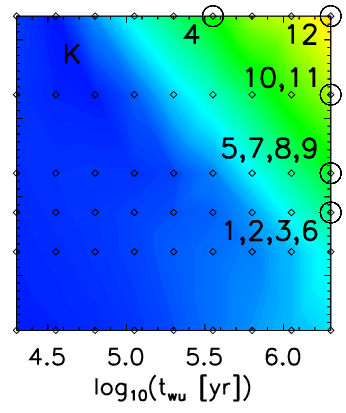}
  \includegraphics[trim={0.22cm 0.1cm 0.07cm 0}, clip, height=3.6cm]{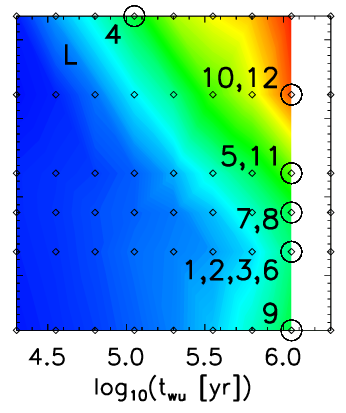}
  \includegraphics[trim={0.22cm 0.1cm 0.2cm 0}, clip, height=3.6cm]{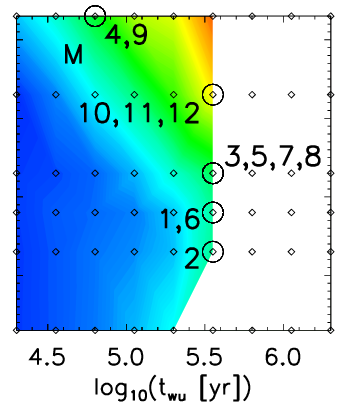}
  \includegraphics[trim={0.1cm 0.1cm 0.6cm 0}, clip, height=3.6cm]{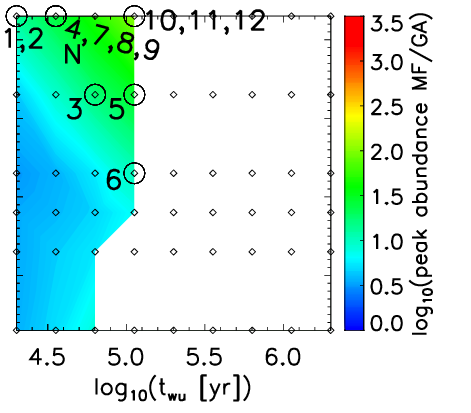}\\
\caption{The ratio of the peak abundances of methyl formate to glycolaldehyde produced by the entire chemical model grid. Each panel corresponds to a specific cosmic-ray ionization rate, $\zeta$ and initial visual extinction, A$_{\text{v,init}}$. Within each panel, models varying in final gas density and warm-up timescale are indicated with diamonds. The ``standard'' model is also marked with a square. The best-matching models for the sources in NGC 6334I, using matching procedure \#2, are marked with circles, with numbers indicating the ID number of the source as listed in Table~\ref{table2}.\label{fig:MFGA}}
\end{sidewaysfigure}

\begin{sidewaysfigure}
\centering
  \includegraphics[trim={0.3cm 0.1cm 0.07cm 0}, clip, height=3.6cm]{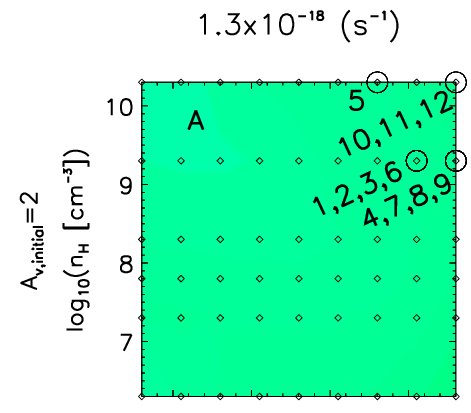}
  \includegraphics[trim={0.22cm 0.1cm 0.07cm 0}, clip, height=3.6cm]{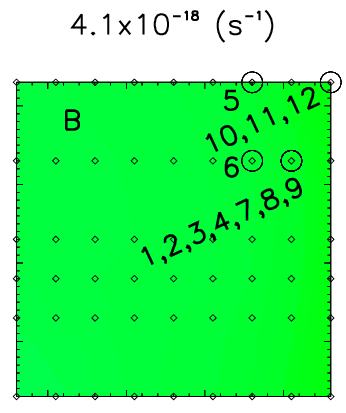}
  \includegraphics[trim={0.22cm 0.1cm 0.07cm 0}, clip, height=3.6cm]{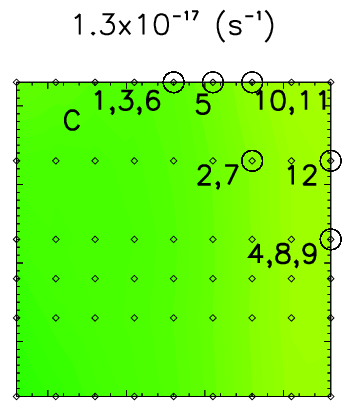}
  \includegraphics[trim={0.22cm 0.1cm 0.07cm 0}, clip, height=3.6cm]{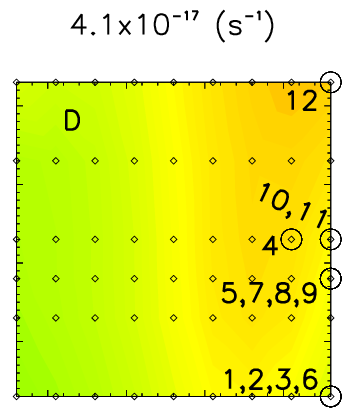}
  \includegraphics[trim={0.22cm 0.1cm 0.07cm 0}, clip, height=3.6cm]{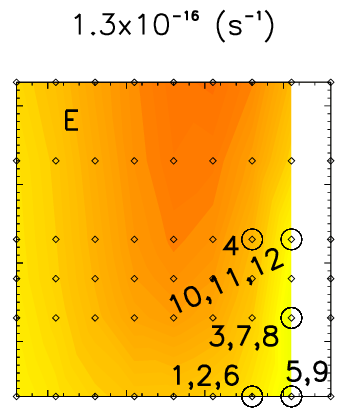}
  \includegraphics[trim={0.22cm 0.1cm 0.2cm 0}, clip, height=3.6cm]{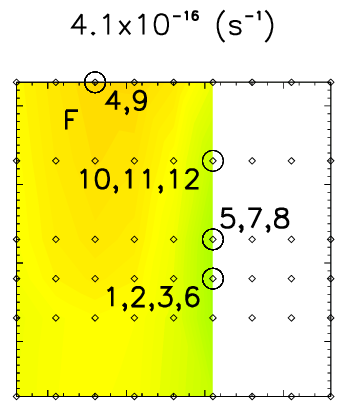}
  \includegraphics[trim={0.1cm 0.1cm 0.6cm 0}, clip, height=3.6cm]{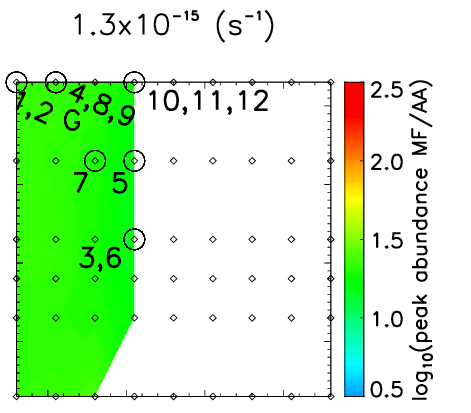}\\
  \includegraphics[trim={0.3cm 0.1cm 0.07cm 0}, clip, height=3.6cm]{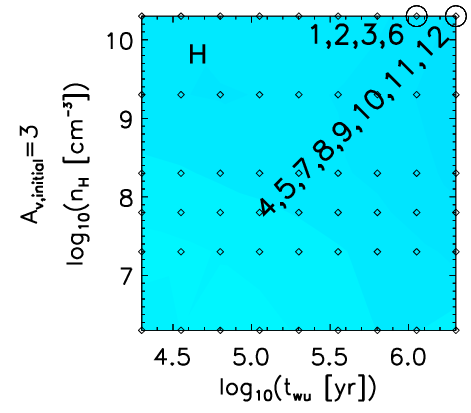}
  \includegraphics[trim={0.22cm 0.1cm 0.07cm 0}, clip, height=3.6cm]{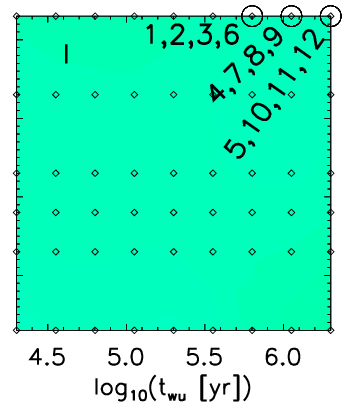}
  \includegraphics[trim={0.22cm 0.1cm 0.07cm 0}, clip, height=3.6cm]{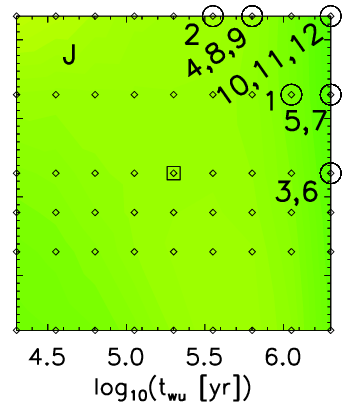}
  \includegraphics[trim={0.22cm 0.1cm 0.07cm 0}, clip, height=3.6cm]{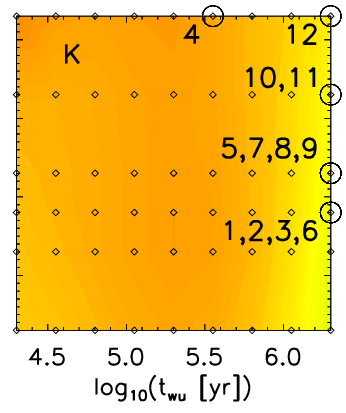}
  \includegraphics[trim={0.22cm 0.1cm 0.07cm 0}, clip, height=3.6cm]{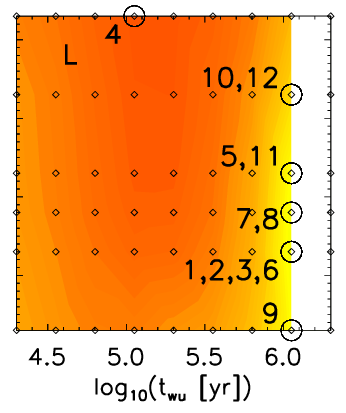}
  \includegraphics[trim={0.22cm 0.1cm 0.2cm 0}, clip, height=3.6cm]{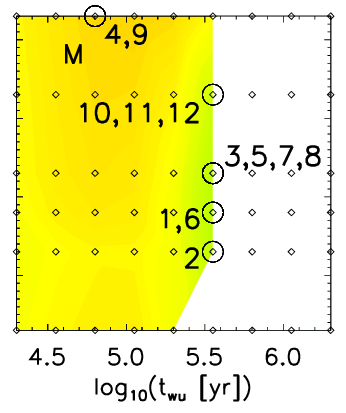}
  \includegraphics[trim={0.1cm 0.1cm 0.6cm 0}, clip, height=3.6cm]{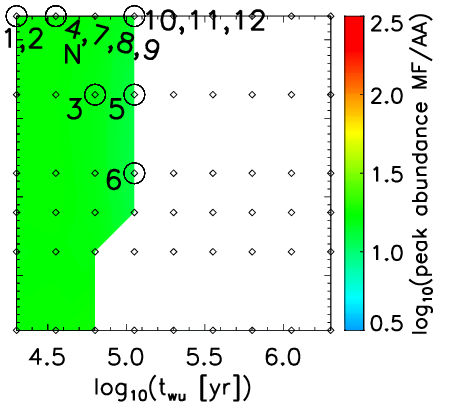}\\
\caption{As Fig.~\ref{fig:MFGA}, but showing the ratio of methyl formate to acetic acid.\label{fig:MFAA}}
\end{sidewaysfigure}

Fig.~\ref{fig:MFGA} maps the logarithm of the MF:GA ratio produced by each model in the parameter space, based on the comparison of peak molecular abundances as described in Sec.~\ref{sec:match}. The panels are labeled A--N, with each panel corresponding to a particular combination of $\zeta$ and A$_{\text{v,init}}$; the upper row corresponds to A$_{\text{v,init}}$=2, and the lower to A$_{\text{v,init}}$=3. Within each panel, the MF:GA ratio is mapped as a color contour as a function of $n_{\text{H}}$ and $t_{\text{wu}}$. The contours are based on the model values at positions marked with diamonds. The highest abundance ratios are shown in red, and the lowest in purple, while regions of white space indicate models that do not meet the abundance thresholds for methanol and MF as described in Section~\ref{sec:ngc}. Fig.~\ref{fig:MFAA} maps the MF:AA ratio in an identical way. In panel J of each figure, the ``standard'' model setup indicated in Table~\ref{table:conds} is marked with a square symbol. Other symbols shown in these figures are described in Sec.~\ref{sec:matching}.

\subsubsection{Cosmic-Ray Ionization Rate, $\zeta$} \label{sec:CR}

When comparing the fourteen contour plots for each of the two molecular ratios, basic trends emerge as to which conditions yield higher abundance ratios. In general, high MF:GA ratios are achieved toward the top-right of each panel (high density and long warm-up timescale). For a fixed $n_{\text{H}}$ and $t_{\text{wu}}$ combination (i.e. a fixed coordinate in any panel), increasing the CRIR (shifting panels from left to right) tends to increase the MF:GA ratio. This occurs as the result of the grain-surface split in MF and GA abundances described in Sec.~\ref{split1}. As noted in Sec.~\ref{general}, higher $\zeta$-values also tend to produce more COMs overall in the ices, especially during the initial collapse stage.

However, the absolute abundances of methanol, MF and GA suffer in the gas phase with the more extreme $\zeta$ values, due to rapid ion-molecule destruction rates. In the extreme cases, this causes those models to produce very low peak COM abundances, hence excluding them from the matching procedures. This is especially true for models with long warm-up timescales, in which the desorption of COMs from the grains in outpaced by their gas-phase destruction. As a result, the highest MF:GA ratios produced by acceptable models are found not for the highest CRIR case but for the next two lower values.

Toward the lower end of the CRIR distribution (left panels), most of the parameter space produces MF:GA ratios that are somewhat less than one order of magnitude, but the MF abundance is always greater than that of GA. The ``standard'' model itself falls into this lower MF:GA ratio regime.

The MF:AA ratio is fairly stable within each panel, but varies slightly for 
%\st{medium to high cosmic ray ionization rates} 
CRIRs of $1.3 \times 10^{-16}$ s$^{-1}$ and higher. The MF:AA ratio is low under conditions of lower CRIR, and increases as the CRIR increases, until the MF:AA ratio drops off again with high CRIR. The fact that the MF:AA ratio is so stable within each panel shows that the final gas density and the warm-up timescale have little influence. Since these are the two model parameters that are most closely associated with the stage-2 chemical behavior, one may infer that the MF:AA ratio is essentially fixed at early times, during the stage-1 collapse. The ``standard'' model (panel J) has a MF:AA ratio of 32, which is in the mid-range of values determined in NGC 6334I.

\subsubsection{Initial Visual Extinction, $A_\text{v,init}$}

The parameter space includes visual extinctions of 2 and 3 mag, which in Figs.~\ref{fig:MFGA} and \ref{fig:MFAA} are arranged as the top seven panels (A--G) and lower seven panels (H--N), respectively. A slight decrease in the abundance ratios of both MF:GA and MF:AA in the 3 mag models in comparison to those at 2 mag suggests that there is only a very slight influence from the visual extinction on the distinct behavior of MF, GA, and AA. In stage 1 of the model, the initial visual extinction determines the degree to which the molecules are shielded from outside UV radiation during the early stages of collapse. All three C$_{2}$H$_{4}$O$_{2}$ isomers are formed strongly in stage 1 of the model.

The slightly higher UV flux seen early in the models with A$_{\text{v,init}}$=2 might, on its own, be expected to lead to somewhat higher photodissociation-driven production of the three isomers in the ice, especially AA. Meanwhile, similar mechanisms might be expected also to destroy those same species to some extent. However, a more important effect related to the visual extinction is the dust temperatures achieved at early times during the collapse. The initial dust temperature in the A$_{\text{v,init}}$=3 case is $\sim$14.7~K, where it remains through much of stage 1, only falling substantially as higher density/extinction conditions (around $n_{\mathrm{H}} > 4 \times 10^3$~cm$^{-3}$ and A$_{\text{v}} > 3.5$) build up toward the end of the collapse. In the A$_{\text{v,init}}$=2 case, the initial dust temperature is $\sim$16~K at early times, and higher dust temperatures than in the A$_{\text{v,init}}$=2 case is maintained through most of the collapse phase. As a result, with the lower A$_\text{v}$, the grain-surface diffusion of hydrogen in particular is faster throughout the stage-1 model, including near the end of the model run, when the gas density is highest.

%%%%%%%%%%%%%%%%%%%%%%%

This has an important impact on grain-surface radical-radical chemistry, which occurs mainly through the nondiffusive 3-B process, with dominant COM production occurring under the low-temperature conditions ($<10$~K) that obtain toward the end of the stage-1 collapse phase. The higher dust temperatures reached in the A$_{\text{v,init}}$=2 case lead to more rapid recombination of radicals with mobile atomic H, reducing the surface coverage of those radicals, thus reducing the likelihood that the spontaneous production of another radical in close proximity would lead to COM production. Conversion of CO to CO$_2$ is also more efficient at higher temperatures, especially above 12~K \citep[see][]{Garrod_2011}, leading to fewer CO-related COMs.

Under low CRIR conditions (panels A and H), while the MF:AA ratio is higher in the A$_\text{v,init}$=2 case (panel A), the overall abundances of MF, GA \& AA are collectively lower due to the weaker production at the highest density conditions. In the A$_\text{v,init}$=3 case (panel H), although the production of AA at early times is weaker due to the lesser UV field, the stronger production of AA late in the collapse stage overcomes this effect, so that ultimately this model has a smaller MF:AA ratio.

In the low CRIR models, the MF:GA ratio also tends to be marginally smaller in the A$_\text{v,init}$=3 case. The difference in MF:AA ratio between the two A$_\text{v}$ values becomes somewhat less important at higher CRIR values, which drives more COM production in the bulk ices throughout the entire stage 1 and stage 2 evolution.

\subsubsection{Hydrogen Number Density, $\text{n}_\text{H}$}

Fig.~\ref{fig:MFGA} shows the most extreme MF:GA ratios values being achieved in high-density, long warm-up timescale models (i.e.~$n_{\mathrm{H}} > 2 \times 10^8$~cm$^{-3}$ and $t_{\text{wu}} > 10^{5}$~yr). As described in Sec.~\ref{split2}, the principal mechanism leading to these extreme values relates to the long lifetimes of GA molecules on the grain surfaces as the ices desorb, compared with those of MF. The latter easily escapes the hot grains as soon as it is exposed on the ice surface, due to its lower binding energy, while GA cannot escape rapidly enough to avoid chemical reactions with atomic H that is abundant in the gas phase. However, this effect only occurs if there is sufficiently high gas density for adsorption of atomic H, and its subsequent reaction with GA, to compete effectively with the thermal desorption of GA. Furthermore, for this process to substantially reduce the GA abundance during desorption, the warm-up timescale must be long enough that the grain-surface GA is not lost to the gas phase too rapidly, giving the atomic H time to destroy it. In each panel of Fig.~\ref{fig:MFGA}, the MF:GA ratio is fairly stable in most of the (density-timescale) parameter space, only rising above a value $\sim$3 when some combination of density and warm-up timescale is achieved (crudely defined as $\log(n_{\mathrm{H}}) 
+ \log(t_{\text{wu}}) \gtrsim 14$).

In Fig.~\ref{fig:MFAA} it may be seen that within each panel, the MF:AA ratio changes little with density, in strong contrast to the MF:GA ratio. The reason for this is that, while AA has an even greater binding energy \citep[6615~K;][]{Burke15} than GA, the H-abstraction reaction with atomic H has a much greater barrier \citep[3710~K in this model, based on Evans-Polanyi fitting; see][]{Garrod_2013} than for its structural isomers, due to the abstraction occurring from a methyl group rather than the aldehyde group present in MF and GA.

\subsubsection{Warm-up Timescale, $t_{\text{wu}}$} \label{sec:timescale}

In Figure \ref{fig:MFGA}, the highest MF:GA ratios tend to concentrate around the medium to long warm-up timescales (i.e.~$t_{\text{wu}} > 10^{5}$~yr). These longer warm-up timescales tend to prolong the gas-phase chemistry occurring in any given temperature range; for MF, GA and AA, that chemistry is primarily destructive. However, a similar effect is seen in prolonging the period over which the grain-surface GA may be attacked by atomic H that adsorbs from the gas phase, preferentially lowering the ultimate peak gas-phase abundance of GA, and thus raising the MF:GA ratio.

Under high CRIR conditions ($> 10^{-16}$~s$^{-1}$), warm-up timescales greater than $\sim$10$^6$~yr can be too destructive to COMs to be a plausible match to observed abundances, with such effects occurring on shorter timescales with increasing CRIR. The white sections in Figs.~\ref{fig:MFGA} and \ref{fig:MFAA} indicate peak MF abundances $<$$10^{-8} \, n_{\text{H}}$ or peak methanol abundances $<$$10^{-6} \, n_{\text{H}}$. This destructive effect is very strongly associated with the degree of ion-molecule destruction occurring in the gas phase during COM desorption.

\subsection{Comparisons with NGC 6334I sources}
\label{sec:matching}

The model grid appears capable of producing broadly appropriate MF:GA and MF:AA ratios for comparison with the observed values on lines of sight toward NGC 6334I. Here we consider more carefully which models match those sources most closely, with a view to understanding the possible physical basis for the observed bimodal behavior.

Table~\ref{table2} lists the results for each matching procedure (see Section \ref{sec:obs}), the observed abundance ratios of MF:GA and MF:AA, and the best-match modeled abundance ratios along with their respective physical conditions. With no constraints on the selection of the best-match models for each of the twelve sources (matching procedure \#1), many of the best-match models tend to have hydrogen number density values on the higher end of the tested range ($2\times10^{10} \text{cm}^{-3}$), with a mix of results for the other parameters. The $\zeta$ values span almost the full three orders of magnitude sampled by the model grid. For most of the best-match models, the MF:GA and MF:AA ratios achieved fall very close to the observed values, within around 10\%. For the observational lines of sight that reach the more extreme MF:GA ratios, the best-match models can reach similarly extreme ratios, but they are not such a close match; for example, the best model for source MM2-iii (ID number 12) reaches MF:GA and MF:AA ratios of 577 and 43.1, respectively, while the observations indicate values 374 and 28.3. It is likely that the lesser quality of match in these cases is due to the limited sampling of parameter space. With a finer resolution in the parameters in the high-density, high-$\zeta$, long warm-up timescale regions of the parameter space, better matches with individual models could probably be found.

Matching procedure \#2 essentially evaluates the best match for each source based on just the models shown within each individual panel in Figs.~\ref{fig:MFGA} \& \ref{fig:MFAA}, corresponding to a particular $\zeta$ and A$_{\text{v,init}}$. Within each panel, the best-match model for each source is marked with a circle and labeled with a source ID number, 1--12, as shown in Table~~\ref{table2}. Sources 1--9 are the MM1 sources; sources 10--12 are the MM2 sources. The best-match model setups shown in the table for procedure \#2 are based on the overall best-matching combination of $\zeta$ and A$_{\text{v,init}}$.

It is apparent that the best matching models in each panel of Figs.~\ref{fig:MFGA} \& \ref{fig:MFAA} tend to lie toward the top-right corner, where the peak abundance of MF:GA ratio is most variable. Additionally, the sources corresponding to the MM2 hot core region (our labels 10,11,12) are consistently grouped together in all panels, and are usually furthest toward the top-right.

The panel in Figs.~\ref{fig:MFGA} \& \ref{fig:MFAA} with the best overall match using procedure \#2 is the top right panel (G), having CRIR of $1.30\times10^{-15}$ s$^{-1}$ and initial visual extinction of 2. The second best matching panel has CRIR of $1.30\times10^{-17}$ s$^{-1}$ and visual extinction of 3 (panel J). The third (C) and fourth (N) best-match panels have similar setups, with CRIR of $1.30\times10^{-17}$ s$^{-1}$ and $1.30\times10^{-15}$ s$^{-1}$, and initial visual extinctions of 2 and 3, respectively.

Thus, while the $\zeta$ and A$_{\text{v,init}}$ can be constrained over the whole list of sources, the top-ranked best-matching {\it panels} are not adjacent in the parameter space. The best-matching panel (G) has an overall $m$-value of 1.050; this would be equivalent to both modeled molecular ratios diverging from the observed values by a factor $\sim$1.64 for all sources. The second best-matching panel has an $m$-value of 1.125, indicating an average divergence by a factor $\sim$1.70. This small difference between models whose $\zeta$-values are divergent by a factor 100 would tend to suggest that the CRIR value is not well constrained by MF/GA/AA ratios at the sampling resolution tested in the grid.

There is also a possibility that MM1 and MM2 sources have different physical conditions (matching procedure \#3). When executing the matching process separately for MM1 and MM2, the best-match models for MM1 remain the same models as for procedure \#2. However, the best match models for MM2 have conditions with a lower cosmic-ray ionization rate of $4.1\times10^{-18}$ s$^{-1}$ (panel B) and a longer warm-up timescale of $2\times10^{6}$ years. Each of the MM2 sources are best matched by the same model. Although the best-match CRIR and warm-up timescale are now different, the high density value of $2\times10^{10}$ cm$^{-3}$ and visual extinction of 2 mag are the same as in matching procedure \#2 for these three sources. This suggests again that gas density is more determinative than CRIR of the final molecular ratios achieved in each model.

%%%%%%%%%%%%%%%%%%%%%%%
%%%%%%%%%%%%%%%%%%%%%%%
%%% DISCUSSION %%%%%%%%
%%%%%%%%%%%%%%%%%%%%%%%
%%%%%%%%%%%%%%%%%%%%%%%
\section{Discussion}

Although the models include a large number of chemical species, understanding how different variations in physical parameters might affect the ratios between the \ce{C2H4O2} isomers methyl formate, glycolaldehyde and acetic acid, is of primary consideration.

The results of the model grid indicate firstly that the range of observational ratios obtained by \citet{El_Abd_2019} can be achieved by the models. Indeed, although we identify the ``best-match'' model for each of the sources in NGC 6334I, a number of different model setups can reproduce the observed values with a reasonable degree of accuracy. For some of the smaller observed MF:GA and MF:AA ratios, some of the models come very close to the exact values (based on values for matching procedure \#1, as seen in Table~\ref{table2}). When the observed values become most extreme, mainly for the MM2 sources, which are based on upper limits for the column density of glycolaldehyde, the models are also capable of reaching appropriately extreme ratios, but they are less likely to be an exact match. This is the result of having only a limited parameter resolution in the model grid. The warm-up timescale parameter has four different values per order of magnitude variation; the cosmic-ray ionization rate two; in most cases, the gas density has only order-of-magnitude variation in the tested values; while visual extinction is assigned only two possible values in total.

One might expect that with a more finely spaced (and computationally expensive) parameter grid, the best-match models might sometimes vary from those found here. For example, while the models produce a (near-)perfect match (using matching procedure \#1) for our source 1 (NGC 6334I MM1-i) in Table~\ref{table2}, the match for source 5 (NGC 6334I MM1-v) is notably less accurate. In the former case, further tuning in the parameters would probably produce only a limited change in the underlying physical parameters. In the latter case, parameter-tuning might produce not only a better match but a substantial change in the underlying parameters due to the degeneracy of the models in reproducing a limited set of observational data.

An example of the potential degeneracy of, in particular, the cosmic-ray ionization rate in producing similar outcomes may be seen for source 11 (NGC 6334I MM2-ii). Matching procedure \#1 by definition provides the overall best-match model from our model grid. Since the observed MF:GA ratio for source 11 is based on an upper limit for glycolaldehyde, any modeled ratio greater than the observed value is deemed a perfect match, and in fact the best-match model produces a MF:GA ratio that is almost double the observed minimum value. Using matching procedure \#2, which requires all sources to adopt a uniform CRIR, the best-match model still appears to be a fairly acceptable match to observations, while the collective CRIR is nearly three orders of magnitude greater than that obtained using matching procedure \#1. When using matching procedure \#3, which allows MM2 sources to adopt their own collective CRIR (versus MM1 sources), the original best-match model returns for source 11. Under this scheme, the other two MM2 sources, 10 \& 12, are also forced to adopt the same CRIR, but the quality of match is not strongly affected in these cases either.

Of the three main physical parameters varied in the grid, CRIR should probably be viewed with the greatest caution as to the importance of its influence over the model MF:GA and MF:AA ratios. The stage-2 gas density and warm-up timescale, however, appear to be somewhat more robust when imposing the three different matching procedures.

\subsection{Comparison of Peak Abundance Values}

As described in Sec.~\ref{sec:match}, the MF:GA ratios are determined based on peak abundance values for each molecule. However, due to their different binding characteristics, there is variation in precise the times and temperature at which MF, GA, and AA peak. In fact, the MF and GA abundances do indeed peak at the same temperature and time in some models
, while in others they technically reach their peaks at very different points. The maximum peak-to-peak temperature difference between MF and GA in all our models is 237~K. In this case, MF and GA both reach something very close to their peak values at around the same temperature, but the MF abundance creeps up marginally over time until the end of the model, creating a seemingly large divergence. An example of this effect may be seen in Figure \ref{fig:abundances1}; GA and MF abundances peak at 242~K and 319~K, respectively, while the actual MF:GA ratio at these two temperatures varies only marginally. A similar effect is seen in Figure \ref{fig:abundances3}. Figure \ref{fig:abundances2} demonstrates a more meaningful difference in peak temperatures of 27~K.

Given the typically large errors on observationally determined excitation temperatures (often tens of K), with which we may associate the peak-abundance temperatures discussed here,
the variations between molecules in the models are at least consistent with observations. For the purposes of comparing between MF:GA ratios within the dataset, the models will necessarily display some variation depending on whether the ratio is taken, for example, at the time/temperature of the MF peak, or that of GA. In the interests of simplicity, this variation is not taken into account in the model comparisons. However, it is unclear to what degree this omission is meaningful without using a more physically accurate physical model of the source structure than the present modeling setup can provide. The observed molecular abundances are dependent on the spatial arrangement of the molecules and the local density and temperature structure. Multi-dimensional hydrodynamical and chemical modeling of hot-cores, coupled with direct simulation of emission spectra, may provide the ultimate solution.

\subsection{Density and timescale thresholds for large MF:GA ratios}

We may tie this varying influence of the physical parameters back to the chemical causes of the variation in the MF:GA ratio as discussed in Secs.~\ref{split1} \& \ref{split2}. It was noted that high $\zeta$-values drive an initial split in the solid-phase abundances of MF and GA, which then carries through to the ultimate gas-phase abundances of those molecules; however, this divergence is limited in scale. The much greater divergence needed to reproduce very large gas-phase MF:GA ratios is the outcome of a very selective grain-surface destruction process affecting GA, that occurs during the period of rapid ice desorption. The long residence time of GA once exposed on the grain surface allows it to be attacked and destroyed by atomic H adsorbed from the gas phase. This effect was seen to occur most strongly in models with high gas density and long warm-up timescales, while an elevated CRIR would tend to increase the amount of H present in the gas to some degree. As found by \citet{Padovani_2018}, the production of atomic H in dense regions is almost exclusively caused by cosmic rays, through interactions with electrons released by the initial H$_2$-ionization event.

%%%%%%%%%%%%%%%%%%%%%%%
\begin{sidewaysfigure*}
\centering
  \includegraphics[trim={0.3cm 0cm 0.2cm 0.25cm}, clip, height=3.62cm]{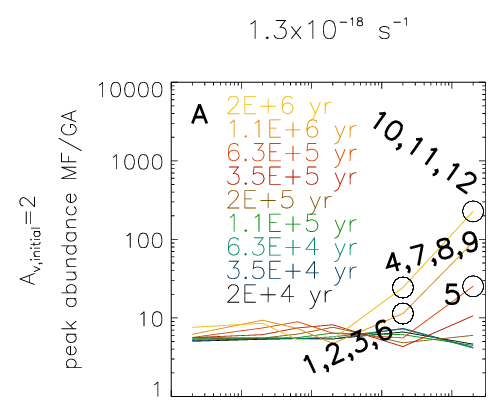}
  \includegraphics[trim={0 0cm 0.2cm 0.25cm}, clip, height=3.62cm]{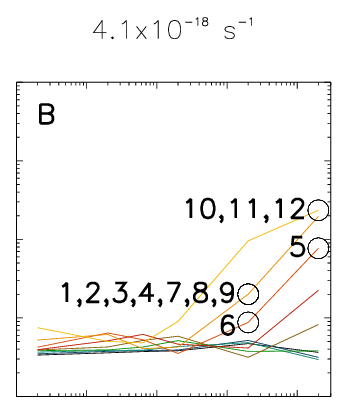}
  \includegraphics[trim={0 0cm 0.2cm 0.25cm}, clip, height=3.62cm]{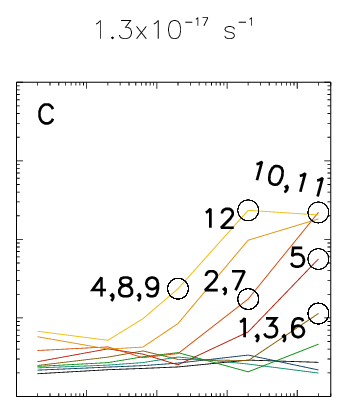}
  \includegraphics[trim={0 0cm 0.2cm 0.25cm}, clip, height=3.62cm]{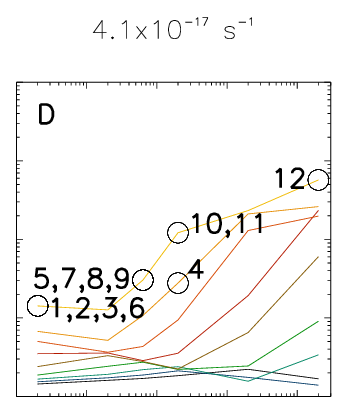}
  \includegraphics[trim={0 0cm 0.2cm 0.25cm}, clip, height=3.62cm]{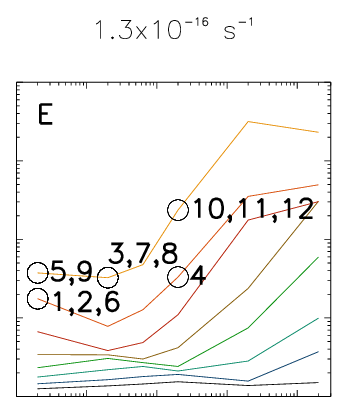}
  \includegraphics[trim={0 0cm 0.2cm 0.25cm}, clip, height=3.62cm]{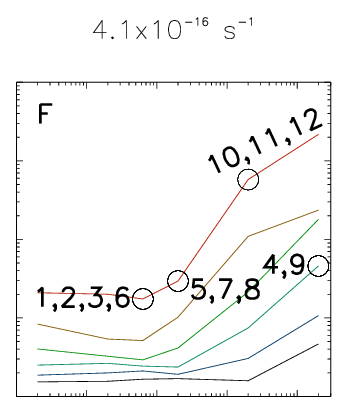}
  \includegraphics[trim={0 0cm 1.36cm 0.25cm}, clip, height=3.62cm]{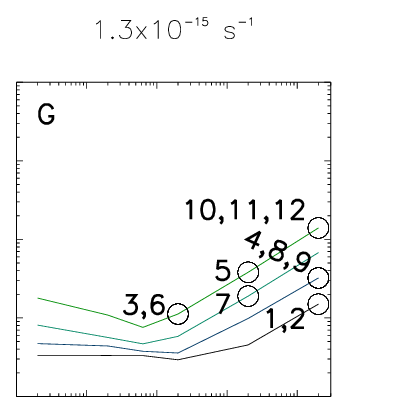}\\
  \includegraphics[trim={0.3cm 0cm 0.2cm 0.25cm}, clip, height=3.62cm]{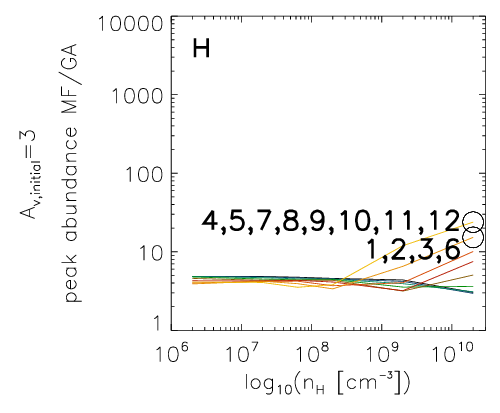}
  \includegraphics[trim={0 0cm 0.2cm 0.25cm}, clip, height=3.62cm]{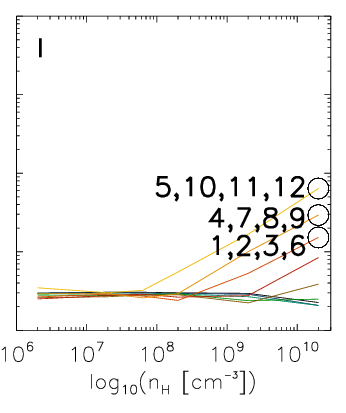}
  \includegraphics[trim={0 0cm 0.2cm 0.25cm}, clip, height=3.62cm]{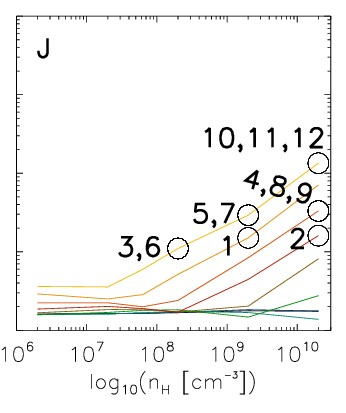}
  \includegraphics[trim={0 0cm 0.2cm 0.25cm}, clip, height=3.62cm]{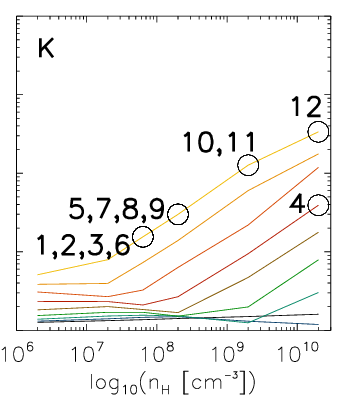}
  \includegraphics[trim={0 0cm 0.2cm 0.25cm}, clip, height=3.62cm]{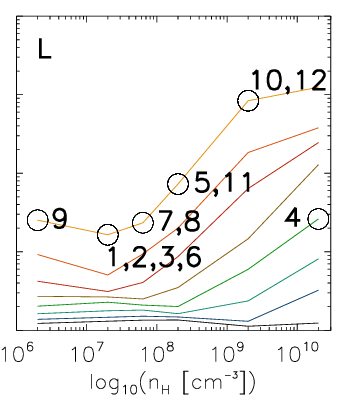}
  \includegraphics[trim={0 0cm 0.2cm 0.25cm}, clip, height=3.62cm]{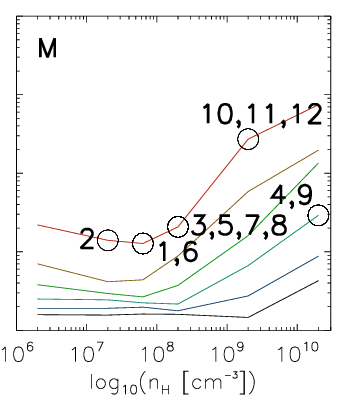}
  \includegraphics[trim={0 0cm 1.36cm 0.25cm}, clip, height=3.62cm]{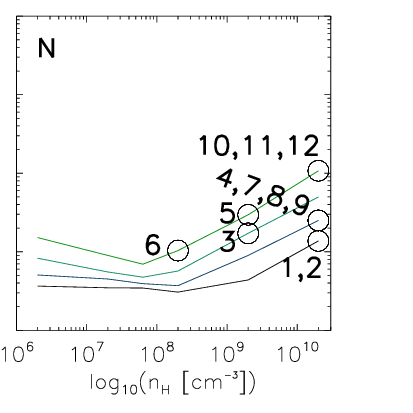}
\caption{Peak MF:GA ratios from the models as a function of stage-2 gas density $n_{\mathrm{H}}$. Each curve corresponds to a different stage-2 warm-up timescales. Best match models from matching procedure \#2 for each panel are circled and marked with their respective source, labeled as per Table~\ref{table2}. Only data points for models that satisfy the minimum abundance thresholds for methanol and methyl formate are shown.\label{fig:thresholds}}
\end{sidewaysfigure*}
%%%%%%%%%%%%%%%%%%%%%%%

To illustrate this effect more clearly, Fig.~\ref{fig:thresholds} plots the peak MF:GA ratio produced in the models against the stage-2 gas density, $n_{\mathrm{H}}$, arranged in panels that correspond to the same fixed values of initial visual extinction and CRIR as in Figs.~\ref{fig:MFGA} \& \ref{fig:MFAA}. Each curve corresponds to a different warm-up timescale, with circles indicating the best-match models for each source under procedure \#2. Regardless of the best-match models in each case, what may be clearly seen in most panels and for most warm-up timescales is the appearance of a threshold gas density above which the MF:GA ratio grows steadily. This is seen most clearly in the lowest-CRIR panels, in which all warm-up timescales achieve very similar MF:GA ratios when gas densities are low. At higher CRIR-values the longer warm-up timescales tend to achieve somewhat higher MF:GA ratios even at low densities. Nevertheless, in all panels a marked uptick is apparent, typically around densities of $\sim$$10^8$ -- $10^9$~cm$^{-3}$, where the MF:GA ratios begin to rise substantially above 10. A more accurate threshold for this rise in MF:GA ratio is given by the product of the gas density and warm-up timescale for each model; a value $n_{\mathrm{H}} t_{\text{wu}} \simeq 3 \times 10^{13}$ cm$^{-3}$yr crudely marks the point at which the upturn begins.

\begin{table}
\centering
	\caption{The time from when GA becomes available to desorb off the grain surface to when it is finished desorbing ($t_{\text{des,GA}}$) is listed next to its respective warm-up timescale.\label{timescales}}
\begin{tabular}{c c}
    \hline
    $t_{\text{warmup}}$ & $t_{\text{des,GA}}$ \\
    (yr) & (yr) \\
    \hline
    $2.00\times10^4$& $1.15\times10^3$\\
    $3.56\times10^4$& $2.05\times10^3$\\
    $6.32\times10^4$& $3.65\times10^3$\\
    $1.12\times10^5$& $6.49\times10^3$\\
    $2.00\times10^5$& $1.15\times10^4$\\
    $3.56\times10^5$& $2.05\times10^4$\\
    $6.32\times10^5$& $3.65\times10^4$\\
    $1.12\times10^6$& $6.49\times10^4$\\
    $2.00\times10^6$& $1.15\times10^5$\\
    \hline
\end{tabular}
\end{table}

While the precise physical conditions most applicable to an individual source are still difficult to determine with certainty based on a comparison using only two key molecular ratios, and while the model grid is only a simplistic representation of the physical conditions in such a source, the models do indicate that extreme MF:GA ratios are the result of some combination of an extreme gas density during the period of rapid ice desorption, and a relatively long period spent in that temperature regime. Below a threshold value of these combined parameters, the MF:GA ratio is expected to be stable at the commonly observed value of $\sim$10, although the precise value is still somewhat dependent on the full set of physical parameters. Considering the critical timescale threshold for adequate COM abundances (see Sec. \ref{sec:timescale}) decreases as the CRIR increases (see Figs.~\ref{fig:MFGA} and \ref{fig:MFAA}), this in turn leads to a shorter timescale needed at high CRIR to reproduce the observed high MF:GA ratio.

Under what specific conditions might this threshold be exceeded? Although average gas densities toward hot cores are often determined to be on the order of $\sim$10$^7$--10$^8$~cm$^{-3}$ \citep{jorg_review2020, herbst_2009}, the presence of a circumstellar disk could plausibly push up the gas densities to rather higher values, such as the maximum adopted in our model grid, of $2 \times 10^{10}$~cm$^{-3}$. In this scenario, the emission from MF, GA, AA and presumably the other COMs detected in the hot core would be originating primarily from that disk (or from its surface, as dust optical depth would likely be important).

It is worth considering the influence of the sticking coefficient of atomic H at these higher temperatures. As in G22, these models assume a sticking coefficient of unity for all species, under all conditions. Measurements by \cite{He_2016} and calculations by \cite{Dupuy_2016} indicate the sticking coefficient of atomic H to be near unity at dust temperatures around 10K, which is appropriate when most of the ice build-up takes place. 
\citep{Dupuy_2016} also calculated rates at dust temperatures as high as 70~K; with a gas temperature of 150~K, they determined a sticking coefficient of H to an amorphous water-ice surface of $\sim$0.3. In our model, GA destruction by adsorbed atomic H occurs mainly in the temperature range $\sim$125--160~K; a decrease in the H adsorption rate by a factor of $\sim$3 in that regime would have a similar effect as lowering the gas-density by that same factor. The use of a temperature-dependent sticking coefficient in the model would therefore be expected to shift the density threshold to a higher value than the present models suggest, likely by a factor of a few.

In accordance with G22, this work assigns a binding energy of 661K to atomic hydrogen \citep{SENEVIRATHNE201759} rather than the value of 450~K that was used in past models \citep{garrod_herbst_2006}. We may consider the effect of this change on the MF:GA ratios; firstly, the greater binding energy produces a lower desorption rate for atomic H. Because desorption is the dominant loss process for surface hydrogen (far beyond any surface reaction) at the high temperatures ($>$100~K) at which surface GA is being destroyed, this increases the surface lifetime of H atoms, which is equivalent to increasing the average surface population of H. This in turn increases the rate of GA destruction by H by a commensurate factor, as the latter is given more time to find its reaction partner before it desorbs. At 150~K, the higher binding energy would provide a surface lifetime around 4 times higher, and thus 4 times more rapid destruction of surface GA. However, this assumes the same diffusion barrier to be used in either case. If the diffusion barrier were scaled down by the same factor as the binding energy, the effect would be lessened (to a net factor $\sim$2). Our use of binding-energy and diffusion-barrier values that were calculated in the same study has the advantage of self-consistency between the two values.

Achieving the highest MF:GA ratios is assisted by elevated $\zeta$-values, due to cosmic rays producing the H atoms in the gas phase that attack GA on the ice surfaces during the desorption period. But the warm-up timescales required under high-$\zeta$ conditions are still on the higher end of the parameter space tested here. How plausible is a warm-up timescale as long as $2 \times 10^6$ yr for a high-mass star-forming source? In fact, such a long timescale is not required to produce the extreme MF:GA ratios. The highly prescribed warm-up profiles used in the present models cover a very wide range of temperatures, not all of which are important to the MF:GA ratio.

At this point it is worth noting more precisely how the nominal warm-up timescale used in the models affects the MF:GA ratio. Due to the nature of the destruction mechanism, the timescale that is important is not the warm-up timescale {\it per se}, i.e.~the period taken to go from 8~K to 200~K; instead, the important timescale is the time taken between the onset of water-ice desorption (when GA trapped in the ice first becomes exposed on the ice surface) and the moment when most GA has left the grains and can therefore no longer be attacked on the grains. This corresponds broadly to a temperature range of $\sim$125--160~K, although this varies somewhat between different models.

Hence, all of the time required to approach the point of onset of water-ice desorption, i.e. the time to go from 8--125~K, should be of little influence to the MF:GA ratio if the MF and GA behavior depends principally on the proposed high-temperature surface-destruction mechanism for GA. Therefore the preference for longer warm-up timescale models in matching the more extreme MF:GA ratios is a constraint on a relatively short window in time, rather than a constraint on the full timescale for hot-core evolution.

Table~\ref{timescales} indicates the time-period taken to pass through the $\sim$125--160~K desorption regime for each of the model warm-up timescales. This desorption timescale is shorter by a factor of $\sim$17 compared with the corresponding full warm-up timescale; all are $\lesssim$10$^5$~yr. The combination of a gas density even higher than the maximum of $2 \times 10^{10}$ cm$^{-3}$ used here, representing gas in a circumstellar disk, combined with an intermediate timescale for desorption on the order of 10$^4$~yr would seem achievable even for rapidly evolving hot-core sources. The attainment of a high MF:GA ratio in such an environment might then be the result of an exceptionally high-density or long-lived circumstellar disk, especially a disk that remained at a temperature of between 125--160~K for a long period.

The timescales to go from 125--160~K may be compared with those determined in the chemical/dynamical models of \citet{Barger_2021}; those authors presented 1-D radiation hydrodynamical (RHD) models for hot cores of several different mass-accretion rates. Chemical models (without nondiffusive dust-grain processes) were then run in post-processing for trajectories traced by Lagrangian particles. Many of these particles remained in the 125--160~K regime for around 1500 yr, although the temperature was not always a monotonically increasing value (see below). Those 1-D models therefore provide a foundation for the timescales shown in Table~\ref{timescales}. Higher-dimensionality RHD models might allow more sustained periods in the appropriate temperature regime with high gas density.

It is worth noting finally that the models involve a monotonically rising temperature with time. The model timescales taken to go from the onset of water-ice desorption to the end of GA desorption are tied to that gradual ramp-up behavior. A model that topped out at a temperature within the 125--160~K range should still ultimately lose all of its water and GA ice (as well as MF and AA) to the gas phase, but the timescale required for complete desorption of the ices would depend on the final value of that temperature. Identifying a very specific timescale or density required to produce a high MF:GA ratio would be best achieved using a more physically and dynamically exact model of hot-core evolution, such as one that includes circumstellar disk evolution.

\begin{table}
\centering
	\caption{Original and reduced activation-energy barrier values for grain-surface/ice reactions of atomic H with glycolaldehyde. Barriers in the test models are reduced by 25\%.\label{acts}}
 \fontsize{8pt}{8pt}\selectfont
 \begin{tabular}{l c c}
 \hline
	Reaction & Original & Reduced \\
             & (K) & (K) \\
 \hline
  \ce{H} + \ce{CH2OHCHO} $\rightarrow$ \ce{H2} + \ce{OH2CCHO} & 992 & 744 \\
		\ce{H} + \ce{CH2OHCHO} $\rightarrow$ \ce{H2} + \ce{CHOHCHO} & 1530 & 1150 \\
		\ce{H} + \ce{CH2OHCHO} $\rightarrow$ \ce{H2} + \ce{CH2OHCO} & 374 & 281 \\
		\ce{H} + \ce{CH2OHCHO} $\rightarrow$ \ce{CH2OHCH2O} & 1100 & 825 \\
		\ce{H} + \ce{CH2OHCHO} $\rightarrow$ \ce{CH2OHCHOH} & 2520 & 1890 \\
		\hline
  \end{tabular}
\end{table}

\subsection{Activation-Energy Barrier Testing}

Having identified a very specific chemical behavior as the cause of the observed extreme MF:GA ratios, it is valuable to test the sensitivity of the model outcomes to the activation-energy barrier adopted for the critical destruction reactions between diffusive atomic H and glycolaldehyde on the grain surfaces. In particular, we have sought to test the effect on the threshold densities beyond which the surface GA mechanism begins to take hold. 

In a set of test models, the activation-energy barriers of the grain-surface/ice reactions of atomic H with glycolaldehyde were reduced by 25\%, as shown in Table~\ref{acts}. This value was chosen to correspond to the approximate degree of variation in activation energy barriers for these reactions calculated by \cite{barcia_2018} when adopting different functionals (see their table A.2).

Three timescales were chosen to evaluate the effect of the lowered reaction barrier, using a setup that adopts the standard $\zeta$-value and an initial $A_{\mathrm{V}}$ of 2. The first timescale chosen ($2\times10^{5}$ yr) is that in which the uptick in MF:GA first begins to manifest strongly in Fig.~\ref{fig:thresholds}.

\begin{figure}
\centering
  \includegraphics[width=\columnwidth]{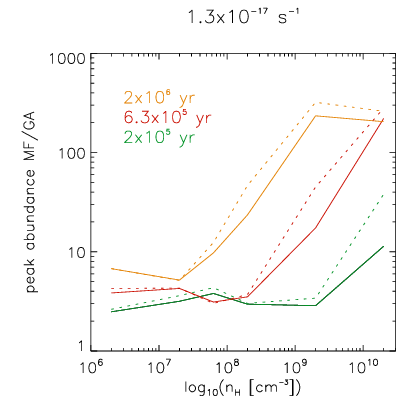}
\caption{A test on the effect of a lowered reaction barrier on GA formation reactions on the peak abundances of MF:GA. Original reaction barrier models are shown as solid lines and barriers reduced by 25\% are shown as dashed lines.\label{fig:lowbarrier}}
\end{figure}

Fig.~\ref{fig:lowbarrier} contrasts the peak MF:GA ratios obtained in the regular models (solid lines) with the new runs (dashed). The lowered reaction barrier data is shifted up slightly in its peak MF:GA ratio. By lowering the barriers for reactions that form GA, those reactions are able to proceed easier and therefore more GA is made in the models. This alteration of the GA reaction barriers has a modest and fairly uniform effect on the models tested, especially at densities greater than the threshold value. However, the threshold density itself is essentially unaffected by the change. We therefore expect that uncertainties in the precise activation-energy barrier value may have a moderate influence on model outcomes, but not on the qualitative result nor on the quantitative determination of the density threshold value.

%%%%%%%%%%%%%%%%%%%%%%%
%%%%%%%%%%%%%%%%%%%%%%%

\subsection{Cosmic-Ray Ionization Rates}

While the cosmic-ray ionization rate is meant to represent a generic galactic value, it is commonly assumed in chemical modeling studies that the rate is constant in a local region. However, there is evidence \citep{Gaches_2018} that the protostar itself can contribute to the cosmic-ray ionization rate especially in denser regions. MM1 would seem to have much more potential for internally accelerated cosmic rays as it is driving a non-thermal jet component \citep{Brogan_2018} and this could be evidence of cosmic-ray acceleration (see for example \cite{Padovani_2021}). In constrast, MM2 shows only thermal radio emission consistent with a hypercompact HII region.

The MM1 sources in NGC 6334I have lower MF:GA abundance ratios than MM2, while our matching procedure \#3 indicates that MM1 sources should collectively be best matched by a much higher CRIR than MM2; a value of $1.30 \times 10^{-15}$ versus $4.11 \times 10^{-18}$~s$^{-1}$.

At the end of stage 1 (collapse), the difference in chemical abundances of our key species, MF, GA, and AA, in our models is mainly affected by the CRIR value. When comparing Figures \ref{fig:abundances1}, \ref{fig:abundances2}, and \ref{fig:abundances3}, the starting abundances of these species is higher in conditions with high CRIR, due to the greater abundance of radicals on grain surfaces and in the ices, as described in Section \ref{general}. Although the CRIR does have an effect on the inheritance of MF, GA, and AA from stage 1 to stage 2, the main influence on high MF/GA ratios appears later on in stage 2, when GA is destroyed on the grain surface as described in ~\ref{split2}.

Given the relatively small effect that the CRIR has (versus that of the gas density and warm-up timescale) on the MF:GA and MF:AA ratios in the models, the apparent variations between MM1 and MM2 may be not ultimately be very meaningful. The large range of $\zeta$-values provided by the unrestricted matching procedure \#1, which notably produces a wide range of values even within the MM1 and MM2 sources, appears to indicate that $\zeta$ is not well constrained.

It is also noteworthy that, in spite of the MM2 sources having the greatest MF:GA ratios (i.e.~only upper limits for the observed GA column densities), these sources are best matched by models with very low CRIR values, which would tend to {\it decrease} the MF:GA ratio in each model. It seems most likely, therefore, that the overall smaller effect of CRIR on the MF:GA ratio is acting to fine-tune the match parameter to produce the best possible match to the observed ratios while not being the main influence on it. Thus, as already noted, the use of a model grid with greater resolution in all of the main physical parameters may be required to tune the CRIR rate in a more meaningful way.

Ultimately, the smaller influence of CRIR on the MF:GA and MF:AA ratios means that these ratios are not sufficient to determine the most likely CRIR for each source. With this particular goal in mind, a larger range of molecules would be necessary for the comparison, such as we have for the comparison of the models with IRAS 16293B and Sgr B2(N2) molecular abundances (see Appendix \ref{appendix}).

\subsection{Influence of luminosity outbursts}

NGC 6334I is one of few (high-mass) star-forming regions found to have exhibited an outburst event; \citet{H17} observed an outburst at mm wavelengths originating in NGC6334 I-MM1 corresponding to a four-fold increase in the millimeter continuum flux density and an estimated 70-fold increase in luminosity near the peak of the event. \citet{hunter_2021} further observed the outburst at mid-IR wavelengths four years into the event, confirming a luminosity increase at a factor of 16.3$\pm$4.4, and noting that the outburst had persisted for at least 6 yr. \citet{H17} noted also that sources MM2--MM4 appear not to have altered in luminosity.

It is interesting to consider the possible influence of an accretion outburst on the scheme described above to explain extreme MF:GA ratios. A luminosity outburst would produce an increase in dust temperature that could either lead to rapid desorption of ice mantles, or indeed, depending on radial distance from the protostar, could elevate colder grains to a temperature conducive to GA destruction. We note that the observational study of the HMYSO G24.33 by \cite{tomoya_2022} shows evidence of changes in thermal line emission due to an accretion event (although in this case it is one of the smaller, periodic events rather than a stronger, stochastic event). In the case of the loss of ice mantles, subsequent cooling of the grains could also lead to the re-adsorption of gas-phase molecules onto the grains. It is therefore difficult to make predictions as the likely effects on the chemical models without an explicit multi-dimensional physical treatments that includes recurring outburst events. Although variable protostellar UV emission is also to be expected, the influence on the large-scale chemistry may be more muted due to local absorption close to the protostar.

We note that, while MM1 has experienced an outburst, extreme MF:GA ratios are thus far detected only toward MM2, which appears stable in its luminosity on recent timescales. Any immediate influence on the abundances of COMs in MM1 might be expected to have shown up in the data of \citet{El_Abd_2019}. However, according to the simulations of \citet{meyer_21}, such outburst events are a common feature of high-mass star formation, accounting for 40--60\% of mass accreted, over multiple events. Those models indicate that individual outbursts may last a few years up to some tens of years, while the integrated burst duration time over the star's lifetime could last up to some hundreds of years. In that case, the monotonically increasing temperature profiles used in the present chemical models may not be accurate. However, the overall time spent in the key temperature range for grain-surface GA destruction may still be valid, especially if the re-accretion of desorbed molecules onto cooling grains can occur.

\section{Conclusions}

We have presented a large parameter grid using the astrochemical model MAGICKAL to investigate the physical conditions that lead to a high methyl formate (MF) to glycolaldehyde (GA) gas-phase abundance ratio in the MM2 hot-core cluster in NGC 6334I observed by \citet{El_Abd_2019}. The present study uses essentially the same model setup as \citet{Garrod_2022} for simulating hot-core chemistry, but with a much larger parameter space tested, focusing mainly on the cosmic-ray ionization rate, final gas density, and warm-up timescale. The effect of varying the initial visual extinction used in the collapse stage of the models was also tested in a limited way. In these models, it was already found that most of the MF and GA, along with many other oxygen- and nitrogen-bearing COMs, were formed on the dust grains, often during the early, cold, collapse stage. Those COMs would be retained in the icy mantles of the dust grains until much higher temperatures were achieved due to the onset of star formation, leading to the desorption of the mantles into the gas phase.

We identify the destruction of GA on dust grains as the most influential cause of the divergence in gas-phase MF:GA ratios between different sources. The mechanism occurs when water-ice desorption becomes rapid at temperatures greater than $\sim$125~K, leading to GA and other COMs in the bulk ice being exposed on the surface of the ice before they desorb into the gas phase. Under the high gas densities present in the hot core, H atoms from the gas phase can adsorb briefly onto the ice surface, where they may react with and destroy molecules on the surface. Methyl formate, the structural isomer of GA, may be attacked by atomic H; however, because it has a relatively low binding energy to a water-ice surface (lower than that of water itself), it actually leaves the surface almost immediately once it is exposed by the loss of the water molecules that would otherwise trap it in the bulk ice. This means that surface reactions of MF with atomic H are rare. Glycolaldehyde, on the other hand, has a high binding energy and remains on the ice surface for a longer period of time before desorbing. Hence, during this period, there are more chances for it to be attacked and destroyed by atomic H coming from the gas. Although acetic acid (AA) also has a large binding energy, leaving it exposed on the surface for a long period, the absence of an aldehyde group (-CHO) on this molecule makes it less liable to attack by atomic H, so the MF:AA ratio is unaffected.

The influence of this mechanism is strongly dependent on the period of time over which the GA molecules are exposed on the icy grain surface prior to desorption, as well as on the number density of atomic H in the gas that may adsorb and attack it. Below a threshold value of the product of gas density and exposure timescale, this chemical attack does not strongly affect the abundance of GA on the grains, meaning that its ultimate gas-phase abundance is also unaffected. Under these conditions the models retain fairly regular MF:GA ratios over a range of physical conditions. Above the density--timescale threshold, GA destruction becomes stronger and stronger, leading to increasingly extreme MF:GA ratios.

We therefore propose that the existence of certain, selected sources with extreme MF:GA ratios, i.e.~extremely low GA abundances, may be the result of the emission from those molecules (and other complex organic molecules, COMs) emanating from a region with (i) an unusually high gas density and/or (ii) an extended period in the critical temperature regime of $\sim$125--160~K at which GA can be attacked while exposed on the icy grain surfaces. This might indicate that the COM emission in those extreme sources is especially dominated by molecules in a circumstellar disk, where those higher densities might be achieved. For sources with more typical MF:GA ratios, emission may be coming from some more extended hot region of lower gas density. A more dynamically exact physical model of hot core evolution to accompany the chemical models would be necessary to explore this idea.

Another important mechanism that produces a divergence in MF and GA abundances occurs in the ice mantles, and becomes influential when cosmic-ray ionization rates are high. The greater dissociation of molecules such as water in the ice mantles, caused by the secondary UV field induced by CRs, leads to a higher production of atomic H in the ice. Above around 17~K, this H becomes highly mobile in the ices, leading to: (i) recombination with large radicals formed earlier in the model, to produce stable molecules including MF; and, (ii) reaction with GA to produce the related ethylene glycol and glyoxal molecules. Although this effect is notable, it does not appear sufficient on its own to reproduce the observed extreme MF:GA ratios.  

The cosmic-ray ionization rate also influences the abundance of atomic H in the gas-phase, which can act to increase the efficiency of GA destruction on the ice surfaces. However, it appears to have a weaker influence on MF:GA ratio than the gas density or warm-up timescale used in the model grid. In models with the most extreme cosmic-ray ionization rates, gas-phase ion-molecule reactions lead to low COM abundances, so that -- regardless of the MF:GA -- the absolute abundances of these molecules are simply too low to be applicable to hot core observations. Models with the highest CRIR values in the parameter space do require a shorter warm-up timescale to reproduce observed MF:GA ratios. Otherwise, long timescales at high CRIR will lead to absolute COM abundances that are too low.

Using a selection of matching parameters to compare the model data with observed molecular ratios for hot core sources in NGC 6334I, the gas density and warm-up timescale used in the models seem to be the most consistently determined parameters; however, the CRIR is not very well constrained. A grid of models with higher resolution in the key physical parameters might improve this situation, but it is more likely that there is simply not enough observational data in the comparison with two observational ratios to allow other than the most powerful influences on MF and GA abundances to be constrained. As a result, we advise some caution in using the best-match chemical models as an explicit guide to the local conditions present in the NGC 6334I sources with which we have done our most extensive comparisons. The MM2 sources appear to be best represented by higher densities in the hot-core emission regions, with a more sustained period of ice desorption rather than a rapid heating period to eject COMs from the grains. These conditions might be complemented by somewhat elevated CRIR values, but this cannot be determined with certainty.

We summarize the main conclusions of this modeling study below:

\begin{enumerate}[noitemsep, label=(\roman*)]
    \item The chemical models are capable of reproducing the full range of MF:GA and MF:AA ratios observed.
    \item A combination of higher gas density and longer timescales for ice desorption appears to be the physical cause of the unusually high MF:GA ratios in the NGC 6334I MM2 sources.
    \item Such conditions might be indicative of COM emission emanating mainly from a dense and perhaps long-lived disk.
    \item The main chemical mechanism driving GA destruction occurs on the icy surfaces of dust grains, when atomic H adsorbed from the gas phase attacks GA before it has time to desorb.
    \item High gas densities allow destructive atomic H to adsorb and attack GA more readily.
    \item A long period of time spent in the temperature range of $\sim$125--160~K (i.e.~between the times when water begins to desorb and when GA has left the grains entirely) leads to more GA destruction by atomic H.
    \item The different binding energies of MF and GA on water ice are crucial to the selectivity of the surface destruction mechanism; individual MF molecules escape the surface as soon as they are exposed by water loss, while GA lingers and can be substantially destroyed by H, given the right conditions.
    \item Below a threshold combination of gas density and warm-up timescale, modeled MF and GA ratios are stable and in keeping with more typical observational values.
    \item Acetic acid on the ice surface is not affected by the presence or absence of atomic H adsorbed from the gas, in spite of its high binding energy, due to the higher barriers to reaction with atomic H.
    \item An increased cosmic-ray ionization rate results in greater production of most COMs in the ice, although extreme values also act to destroy gas-phase COMs very rapidly. A higher CRIR also produces modest enhancement in the MF:GA ratio, but not sufficient to explain the observed extreme MF:GA ratios.
    \item Varying the initial visual extinction produces no significant changes in the MF:GA ratio.
\end{enumerate}

\section*{Acknowledgements}

This work was funded by the National Science Foundation, through grants AST 1906489 and AST 2206516.
This paper makes use of the following ALMA data: ADS/JAO.ALMA\#2012.1.00712.S, ADS/JAO.ALMA\#2013.1.00278.S,
 ADS/JAO. ALMA\#2011.0.00017.S, and ADS/JAO.ALMA \#2012.1.00012.S. ALMA is a partnership of ESO (representing its member states), NSF (USA) and NINS (Japan), together with NRC (Canada), NSTC and ASIAA (Taiwan), and KASI (Republic of Korea), in cooperation with the Republic of Chile. The Joint ALMA Observatory is operated by ESO, AUI/NRAO and NAOJ. The National Radio Astronomy Observatory is a facility of the National Science Foundation operated under cooperative agreement by Associated Universities, Inc. We also thank M\'elisse Bonfand for helpful discussions.

\section*{Data Availability}

Chemical model output data will be made available to other researchers upon request.

\appendix

\section{Comparisons with IRAS 16293B and Sgr B2(N)}\label{appendix}

Following G22, we also compare our model results with observations of IRAS 16293B and Sgr B2(N2), using column density data from the PILS~\citep{jorg2016} and EMoCA~\citep{belloche_2016} ALMA line surveys, respectively. The comparison conducted by G22 was based on chemical model results from only three model runs, using different warm-up timescales ({\it fast}, {\it medium} and {\it slow}). All other physical and chemical model parameters were fixed in the G22 model comparison. Abundances in the low-mass star-forming source IRAS 16293B were found to correspond more closely to the {\it fast} warm-up model, while abundances toward the high-mass source Sgr B2(N2) matched more closely with the {\it slow} model. With the larger model grid explored here, we seek to determine whether the warm-up trend holds, and what degree of degeneracy may apply when other parameters are considered.

\subsection{Matching Parameter}

When considering the column densities observed in the PILS and EMoCA surveys, the match parameter includes a selection of observed molecules, thus Eq.~(\ref{eq:match1}) includes multiple terms. For each term, given by Eq.~(\ref{eq:match2}), $R_\text{mod,i}$ represents the ratio of the peak fractional abundance of molecule $i$ versus that of methanol, while $R_\text{obs,i}$ represents the observed column density ratio of the same two molecules.

The observed ratios are the same as those used by G22, which were provided in the review of \citet{jorg_review2020}; for IRAS 16293B, this included a total of 24 molecular column densities from the PILS survey, while for Sgr B2(N2) 19 values were available from the EMoCA survey. G22 compared their model values against the observational data for all of these species, although they did not calculate a matching parameter and instead judged the best-match model (out of three warm-up timescales) based on a qualitative judgement. 

Here, in order to determine the best match model based on the matching parameter, we further exclude some species from the analysis for which the G22 chemical network is uncertain or likely incomplete, namely: acetone, (CH$_3$)$_2$CO; propanal, C$_2$H$_5$CHO; methyl isocyanate, CH$_3$NCO; methyl isocyanide, CH$_3$NC; acetamide, CH$_3$C(O)NH$_2$; and methyl mercaptan, CH$_3$SH (see G22 for a brief discussion of this issue). We also exclude cyanoacetylene, HC$_3$N, due to the fact that its abundance is not exclusively associated with the hot-core chemistry, meaning that a good match is more closely related to having a model with an accurate physical/spatial structure. Therefore, there are a total of 17 molecular column densities considered in the match parameter for the PILS survey (IRAS 19293B) and a total of 13 molecules column densities considered for the EMoCA survey (Sgr B2(N2)).

\subsection{Results} \label{sgr_iras}

 The model that has the lowest $m$-value is deemed the best-match model over all of the parameter space, indicating which physical conditions may be most representative of these particular sources (in the context of using a single-point model to represent the entirety of each source). However, as there can be some flexibility in the conditions that still yield low match parameters, we present the top four best-match models for the PILS and EMoCA results, listed in Table \ref{tab:iras_sgr} along with each match parameter. Additionally the model that is the worst match to the data (has the highest match parameter value) is included in Table \ref{tab:iras_sgr} along with its match parameter for reference. Note that the match parameter for IRAS 16293B is naturally larger than that of Sgr B2(N2), due to the larger number of molecules in the matching procedure (17 versus 13).

For both sources, the best match models have CRIR values around the canonical value, with the best matches for EMoCA trending slightly lower. High CRIR values appear to provide a poor match. The gas densities of the best-match models are a little higher for the EMoCA matches. For both sources the worst-match model has the highest gas density value used in the model grid. Longest warm-up timescale ($2 \times 10^6$ yr) is favored in all the best matches for the EMoCA data, while for PILS the warm-up timescale is somewhat lower (around $10^5$ yr). This broadly agrees with the trend noted by G22 in their limited parameter space. All the best matches for each source have initial visual extinction of 3.

\begin{figure*}
\centering
  \includegraphics[width=3.5in]{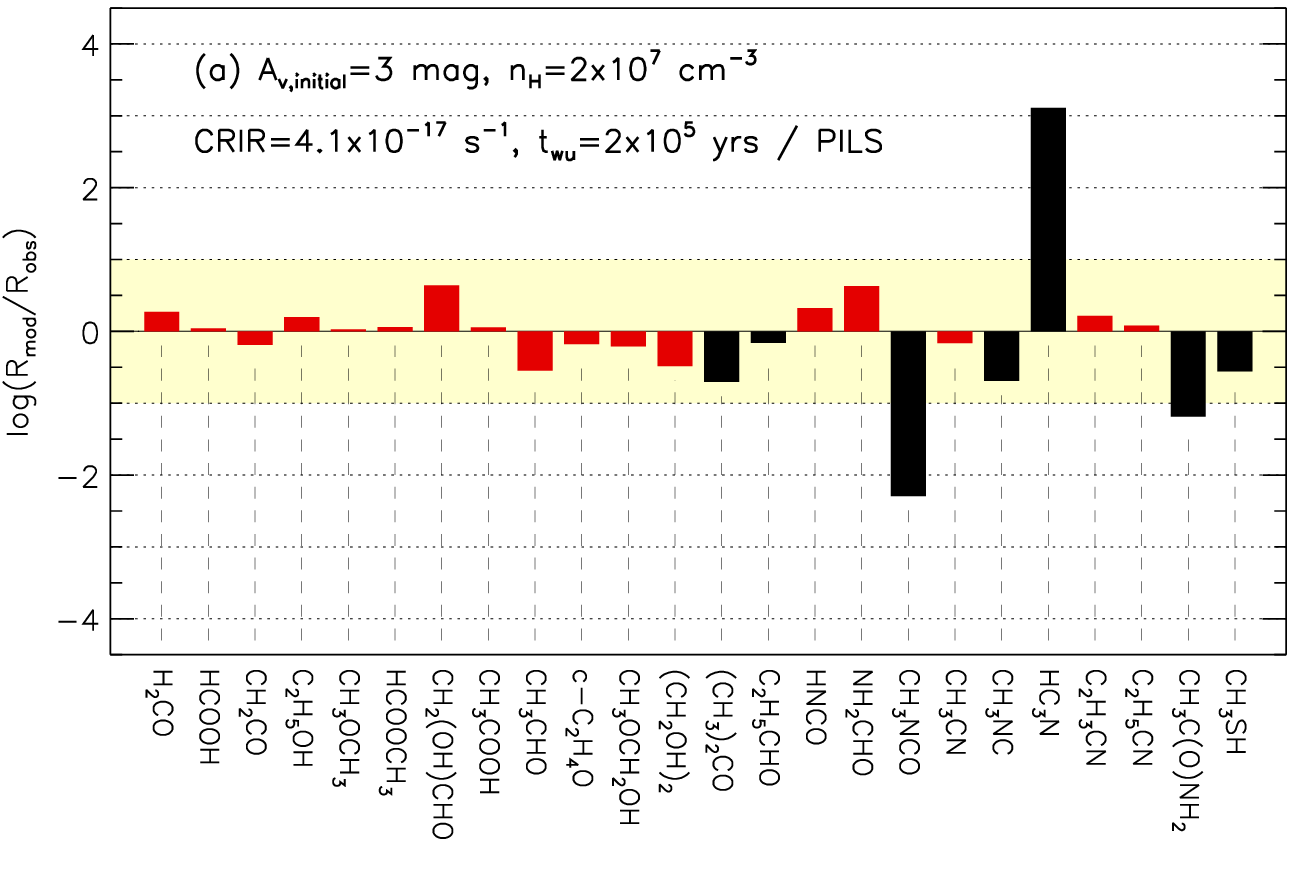}
  \includegraphics[width=3.5in]{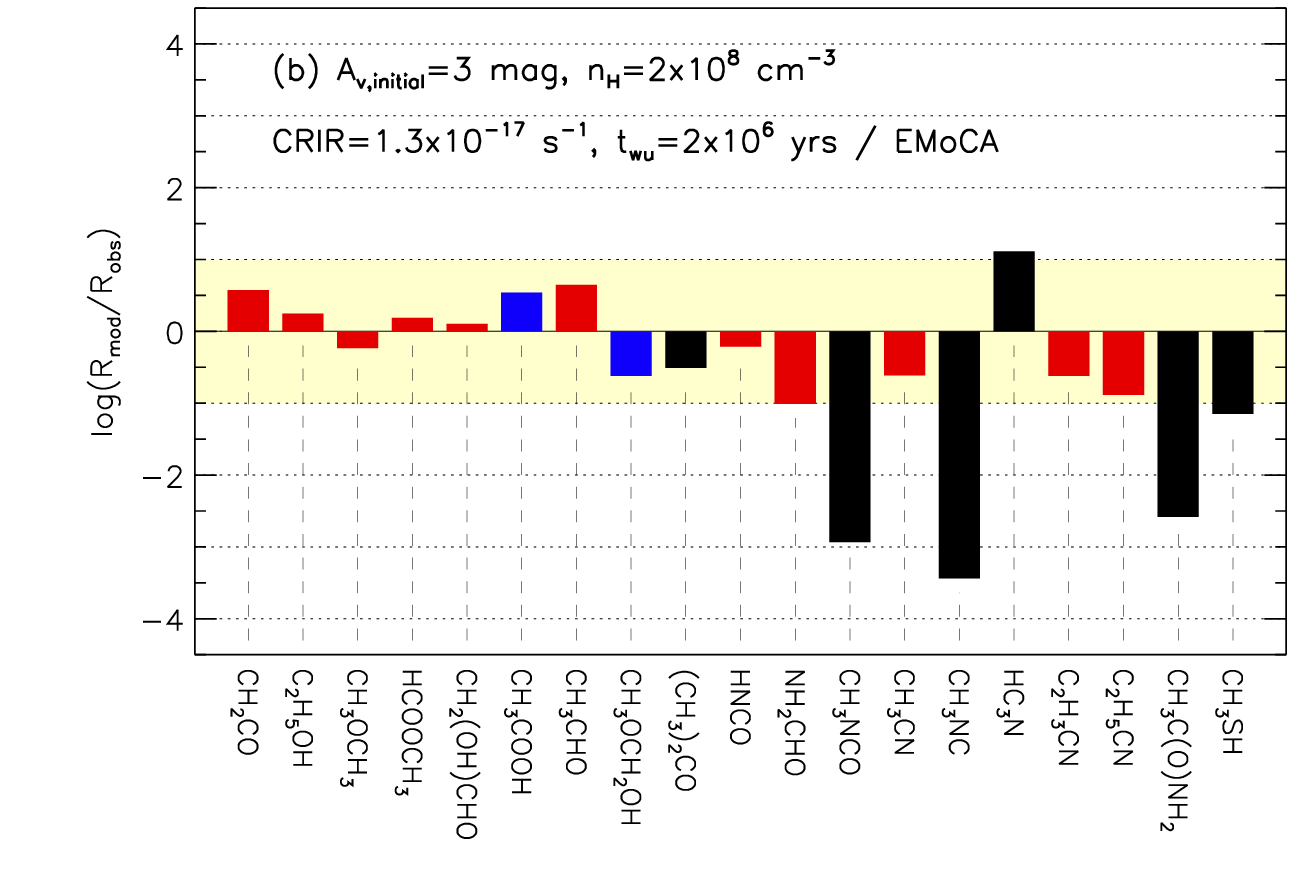}
  \vspace{-2mm}
\caption{Comparison between the peak gas-phase molecular abundances achieved in the best-match models (normalized to the models’ peak methanol abundances), $R_{\mathrm{mod}}$, and observational values of the same quantities (based on column densities), $R_{\mathrm{obs}}$. Panel (a) corresponds to chemical species observed in the PILS survey of IRAS 16293B \citep{jorg2016}; panel (b) corresponds to species observed in the EMoCA survey of Sgr B2(N2) \citep{belloche_2016}.
Bars indicate the logarithm of $R_{\mathrm{mod}}/R_{\mathrm{obs}}$. Blue bars are based on observational upper limits; bars below the zero-line thus represent an exact match. Black bars are not included in the matching procedure, but are included for completeness. The shaded area represents values where the models and observations vary by 1 order of magnitude or less.}
\label{fig:barplot}
\end{figure*}

The individual $m$-values for each molecule in the best-match model for either source are plotted as bars in Fig.~\ref{fig:barplot}. The red and blue bars are molecules considered in the match, while black bars are not considered, due to the difficulty involved in reproducing these species with the current chemical network, using a single-point model. Blue bars in the plot for Sgr B2(N2) represent upper limits for those molecules; bars falling below the zero-line therefore count as an exact match. The top four best match models show very similar peak abundance behavior, so only the best-match model data are shown here.

\begin{table}
\caption{Four best match models for IRAS 16293B (PILS) and Sgr B2(N2) (EMoCA). Best models determined by lowest overall match parameters. Match parameter considers the selection of molecules listed in Figure \ref{fig:barplot}. Model that is the worst match to the data is also included along with its respective match parameter value for reference.\label{tab:iras_sgr}}
\centering
    \fontsize{8pt}{8pt}\selectfont
    \begin{tabular}{m{5mm} c m{5mm} c c c}
    \hline
    & $m$-value & A$_{\text{v,init}}$ & n$_\text{H}$ & $\zeta$ & $t_{\text{wu}}$ \\
    &  &(mag) & (cm$^{-1}$) & (s$^{-1}$) & (yr) \\
    \hline
    PILS& & & & & \\
    $\#$1& 1.33& 3& $2.00\times10^{7}$&  $4.11\times10^{-17}$& $2.00\times10^5$\\
    $\#$2& 1.40& 3& $2.00\times10^{6}$&  $4.11\times10^{-17}$& $2.00\times10^5$\\
    $\#$3& 1.44& 3& $6.32\times10^{7}$&  $1.30\times10^{-16}$& $6.32\times10^4$\\
    $\#$4& 1.47& 3& $2.00\times10^{7}$&  $4.11\times10^{-17}$& $1.12\times10^5$\\
    $\#$756& 10.92& 2& $2.00\times10^{10}$&  $4.11\times10^{-16}$& $2.00\times10^6$\\
    \\
    EMoCA& & & & & \\
    $\#$1& 1.95& 3& $2.00\times10^{8}$&  $1.30\times10^{-17}$& $2.00\times10^6$\\
    $\#$2& 1.97& 3& $2.00\times10^{9}$&  $4.11\times10^{-18}$& $2.00\times10^6$\\
    $\#$3& 1.98& 3& $6.32\times10^{7}$&  $1.30\times10^{-17}$& $2.00\times10^6$\\
    $\#$4& 1.99& 3& $2.00\times10^{8}$&  $4.11\times10^{-18}$& $2.00\times10^6$\\
    $\#$756& 7.84& 2& $2.00\times10^{10}$&  $4.11\times10^{-16}$& $2.00\times10^6$\\
    \hline
    \end{tabular}
\end{table}

%%%%%%%%%%%%%%%%%%%%%%%
%%%%%%%%%%%%%%%%%%%%%%%

In both panels, modeled values for all of the molecules included in the matching procedure come within one order of magnitude (or better) of the observed values. In the IRAS 16293B comparison, the black-bar molecules CH$_3$NCO, HC$_3$N, and CH$_3$C(O)NH$_2$ deviate at least two orders of magnitude from the observations. For Sgr B2(N2), methyl isocyanate (CH$_3$NCO), methyl isocyanide (CH$_3$NC) and acetamide (CH$_3$C(O)NH$_2$) provide a particularly poor match. However, since these species were not included in the matching procedure, it is not a surprise that they are not well represented in the data shown here.

For the PILS data (IRAS 16293B), the top five molecules that match observations very well are formic acid (HCOOH), dimethyl ether (CH$_3$OCH$_3$), methyl formate (HCOOCH$_3$), acetic acid (CH$_3$COOH), and ethyl cyanide (C$_2$H$_5$CN). For the EMoCA data (Sgr B2(N2)), the five molecules that match the observations most accurately are ethanol (C$_2$H$_5$OH), dimethyl ether (CH$_3$OCH$_3$), methyl formate (HCOOCH$_3$), glycolaldehyde (CH$_2$(OH)CHO), and isocyanic acid (HNCO).

\subsection{Discussion and Conclusions}

In G22, out of the three models considered, the fast ($5\times10^4$ yr) warm-up model agreed best with the IRAS 16293B data and the slow ($1\times10^6$ yr) model agreed best with Sgr B2(N2). In the present results (Table \ref{tab:iras_sgr}), we find that an intermediate warm-up timescale model ($2.00\times10^{5}$) agrees best for IRAS 16293 and that a slow model ($2.00\times10^{6}$) still agrees best for Sgr B2(N2). When comparing the results of Fig.~\ref{fig:barplot} with the equivalent bar plots of the ``fast'' model for PILS and the ``slow'' model for EMoCA considered in G22, we see that there are not extreme differences concerning the chemical species predictions. The three models considered in G22 all had initial visual extinctions of 3 mag, final hydrogen number densities of $2.00\times10^{8}$ cm$^{-1}$, and cosmic-ray ionization rates of $1.30\times10^{-17}$ s$^{-1}$, while our own best-match models were not very far from these same values.

From Table~\ref{tab:iras_sgr}, it appears that Sgr B2(N2) has higher density matches than IRAS 16293B data, which is compatible with the fact that IRAS 16293B is a low-mass source and Sgr B2(N2) is a high-mass source. However, it is notable that neither source has an extreme MF:GA ratio, and also that the best-matching density values are generally on the lower end of the modeled range. If an extreme gas density were associated with a circumstellar disk, this might indicate that COM emission in both of these sources is not strongly associated with high-density regions of a disk structure.

The similarity of the top four model results in their predictions of chemical species abundances indicates that there are different setups that can produce acceptable results, even with the larger range of molecules included in the analysis. However, the best matching models for Sgr B2(N2) all have the longest warm-up timescale available in the grid, and the best-matching timescales for IRAS 16293 are all fairly close together. The $\zeta$-values selected for each source are also quite strongly clustered, with the IRAS 16293 best-match model showing a marginally higher value than Sgr B2(N2) ($4.11 \times 10^{-17}$ versus $1.30 \times 10^{-17}$~s$^{-1}$, respectively).

With this larger selection of molecules in the comparison, all of the physical parameters in the grid appear to be consistently constrained. The overall match parameter for IRAS 16293 is actually based on a larger number of molecules, so the smaller $m$-value that is actually achieved for its best-match model indicates a significantly better match than is achieved for Sgr B2(N2). The better quality of match may be seen by inspection of Fig.~\ref{fig:barplot}. We note again that the black bars are not included in the matching procedure because they are not expected to be well reproduced by the present models (see G22 for discussion of this point). Acetone ((CH$_3$)$_2$CO) and propanal (C$_2$H$_5$CHO) are nevertheless quite well reproduced in the present results. We note also that the blue bars, which correspond only to an observational upper limit, are deemed to be a perfect match to observations if the bar falls below the zero-line; methoxymethanol (CH$_3$OCH$_2$OH) therefore seems to be well matched in the models, which was also found by G22.

In summary, we used the large parameter space to re-evaluate the comparison between the previous smaller collection of MAGICKAL models presented in G22 to observational data taken by the PILS~\citep{jorg2016} and EMoCA~\citep{belloche_2016} surveys of IRAS 16293B and Sgr B2 (N2), respectively. For the EMoCA data of Sgr B2(N2), the best match models have higher densities as expected, standard or lower cosmic-ray ionization rates, and uniformly long timescales. For the PILS data of IRAS 16293B, the best match models have lower densities, standard to slightly higher cosmic-ray ionization rates, and medium timescales. Neither set of models falls into the most extreme values of any physical parameter in the grid.

\bibliography{biblio}{}
\bibliographystyle{aasjournal}

\end{document}